\documentclass[reprint,superscriptaddress,amsmath,amssymb,aps,prc]{revtex4-1}

\usepackage{graphicx}
\usepackage{hyperref}
\usepackage{color}
\usepackage[english]{babel}
\usepackage{bm}
\usepackage[T1]{fontenc}

\usepackage[flushleft]{threeparttable}

\usepackage{CJKutf8}

\newcommand{\tj}[6]{ \begin{pmatrix}
   #1 & #2 & #3 \\
   #4 & #5 & #6 
\end{pmatrix}}

\newcommand{\sj}[6]{ \begin{Bmatrix}
   #1 & #2 & #3 \\
   #4 & #5 & #6 
  \end{Bmatrix}}

\begin{document}

\begin{CJK*}{UTF8}{gkai}
\title{Study of the magnetic octupole moment of $^{173}$Yb using collinear laser spectroscopy}
    \author{R.~P.~de Groote}
    \email{ruben.p.degroote@jyu.fi}
    \affiliation{Department of Physics, University of Jyv\"askyl\"a, PB 35(YFL) FIN-40351 Jyv\"askyl\"a, Finland}
    \author{S.~Kujanp\"a\"a}
    \affiliation{Department of Physics, University of Jyv\"askyl\"a, PB 35(YFL) FIN-40351 Jyv\"askyl\"a, Finland}
    \author{\'A. Koszor\'us}
    \affiliation{Department of Physics, University of Liverpool, Liverpool L69 7ZE, United Kingdom}
    \author{J.G.~Li~(李冀光)}
    \affiliation{Institute of Applied Physics and Computational Mathematics, Beijing 100088, China}
    \author{I.~D.~Moore}
    \affiliation{Department of Physics, University of Jyv\"askyl\"a, PB 35(YFL) FIN-40351 Jyv\"askyl\"a, Finland}
    
\begin{abstract}
    The hyperfine constants of the $^3$P$^{\circ}_2$ state in neutral Yb have been measured using three different dipole transitions. This state was recently shown to have a comparatively large hyperfine magnetic octupole splitting, and thus a puzzlingly large magnetic octupole moment. The measurement is performed using collinear laser spectroscopy on a fast atomic beam, which provides a straightforward route to probing long-lived metastable atomic states with high resolution. From the combined analysis of all three lines we find no significant evidence for a non-zero octupole moment in $^{173}$Yb. 
\end{abstract}

\maketitle
\end{CJK*}

\section{Introduction}

For many decades, measurements of the hyperfine structure of ions and atoms have to extract information on the electromagnetic moments of atomic nuclei~\cite{neyens2003,Campbell2016}. Continuous developments in the field of radioactive ion beam production and manipulation, as well as improvements in spectroscopic methods, have resulted in the measurement of magnetic dipole moments $\mu$ and electric quadrupole moments $Q$ of a wide variety of atomic nuclei~\cite{Mertzimekis2016}. However, measurements of the next term in the multipolar expansion of the hyperfine interaction, the magnetic octupole constant $C$ and the corresponding nuclear moment $\Omega$, are exceedingly scarce. To our knowledge, the magnetic octupole splitting has been measured only for about 18 elements \cite{Daly1954,Jaccarino1954,Kusch1957,McDermott1960,olsmats1961,Faust1961,Faust1963,Brown1966,Blachman1967,Unsworth1969,hull1970,Landman1970,Childs1978,Brenner1985,Childs1991,Jin1995,Gerginov2003,Lewty2012,Singh2013}. This lack of data renders interpretation of these octupole moments challenging, and meaningful nuclear-theoretical progress has therefore not been made over the past half-century. 

From the experimental perspective, however, the past few years have demonstrated promise; a measurement of $\Omega$ in $^{133}$Cs was made using precise optical spectroscopy of the D$_2$ line~\cite{Gerginov2003}, while a measurement on trapped barium ions resulted in the most precise value of $\Omega$ to date~\cite{Lewty2012}. Yet another example was reported in \cite{Singh2013}, where the $\Omega$ of Yb was extracted from the hyperfine intervals in the $^3$P$^{\circ}_2$ state of $^{173}$Yb. This latter measurement is of particular interest, since a large octupole splitting was observed, and a correspondingly large value of $\Omega$ was thus obtained. If verified, the effect would be sufficiently large that measurements of $\Omega$ on radioactive ions would be within reach of spectroscopic methods currently in use at radioactive ion beam facilities~\cite{Campbell2016}. In addition, the value of $\Omega$ obtained in~\cite{Singh2013} is considerably larger than a single-particle shell model estimate. This implies a puzzling deficiency in our understanding of the nuclear electromagnetic moments.

Most recently, the singly-charged Yb$^{+}$ ion was suggested to be better suited for the extraction of higher-order nuclear moments such as $\Omega$ using high-precision trapped-ion techniques \cite{xiao2020}. In this work, however, we seek to validate the measurements in \cite{Singh2013} using collinear laser spectroscopy, which is the standard tool for high-resolution optical spectroscopy of radioactive beams. A successful measurement of a non-zero magnetic octupole constant using this method would pave the way for future measurements on radioactive ytterbium isotopes. We hope this work will motivate and support future measurements and theoretical investigations into $\Omega$ of both stable and radioactive isotopes. 

The structure of this article is as follows. First, in section \ref{theory}, we provide the definitions and procedures which were followed to perform atomic structure calculations. Then, in section \ref{experiment}, we describe the experimental apparatus and the data analysis strategy. In section \ref{conclusion}, we discuss our new value of $\Omega$ for $^{173}$Yb. We will show that with our new experimental value and atomic calculations, the nuclear magnetic octupole puzzle in $^{173}$Yb is resolved.

\section{Theory}\label{theory}

\subsection{Definitions}

Nuclear moments can be extracted from measurements of atomic hyperfine structure in a nuclear-model independent way. The hyperfine shift $E_F^{(1)}$ of a state, with angular momentum $F = I + J$ (with $I$ the nuclear spin and $J$ the electronic spin) due to the interaction with the nuclear moments $\mu, Q, \Omega, \ldots$, is given by 
\begin{align}
    E_F^{(1)} = \sum_k M_k(I,J,F) \left\langle II \right| T_k^{(n)}  \left| II \right\rangle \left\langle JJ \right| T_k^{(e)}  \left| JJ \right\rangle,  \label{eq:hyp}
\end{align}
where
\begin{align}
    M_k(I,J,F) = (-1)^{I+J+F}\frac{\sj{I}{J}{F}{J}{I}{k}}{\tj{I}{k}{I}{-I}{0}{I} \tj{J}{k}{J}{-J}{0}{J}}.\notag
\end{align}
Defining $K = F(F+1) - I(I+1) - J(J + 1)$, this can be written as  (truncated at the octupole $(k=3)$ term):
\begin{widetext}
\begin{align}
    E_F^{(1)} & = \frac{AK}{2} + \frac{3B}{4}\frac{K(K+1) - I(I+1)J(J+1)}{(2I(2I-1)J(2J-1))}\notag \\ 
        &  + \frac{5C}{4}\frac{K^3 + 4K^2 +\frac45 K(-3I(I+1)J(J+1) + I(I+1) + J(J+1) + 3) - 4I(I+1)J(J+1)}{I(I-1)(2I-1)J(J-1)(2J-1))},
\end{align}
with hyperfine constants
\begin{eqnarray}
\label{HFScon}
    A & = \frac{1}{IJ} \left\langle II \right| T_2^{(n)}  \left| II \right\rangle \left\langle JJ \right| T_1^{(e)}  \left| JJ \right\rangle & = \frac{\mu_I}{IJ} \left\langle JJ \right| T_1^{(e)}  \left| JJ \right\rangle \notag\\
    B & = 4 \left\langle II \right| T_2^{(n)}  \left| II \right\rangle \left\langle JJ \right| T_2^{(e)}  \left| JJ \right\rangle & = 2eQ  \left\langle JJ \right| T_2^{(e)}  \left| JJ \right\rangle \\
    C & =  \left\langle II \right| T_3^{(n)}  \left| II \right\rangle\left\langle JJ \right| T_3^{(e)}  \left| JJ \right\rangle&  = -\Omega \left\langle JJ \right| T_3^{(e)}  \left| JJ \right\rangle. \notag 
\end{eqnarray}
\end{widetext}
Equation \ref{HFScon} illustrates how extracting the nuclear moment from a measurement of the hyperfine splitting requires either a reference nucleus in which this moment was measured through other means, or atomic-structure calculations to evaluate the electronic matrix elements for an electronic state $|J M_J \rangle$. For $k>2$, i.e. all moments beyond the electric quadrupole moment, only the latter option is currently available. Furthermore, extracting the hyperfine constants from experimental spectra requires evaluation of the second-order shifts due to mixing of close-lying atomic states. The second-order shift $E_F^{(2)}$ of a level $F$ due to mixing with another state $\alpha$ with electronic angular momentum $J_\alpha$ is defined as 
\begin{eqnarray}
E_F^{(2)} &=& \sum_{\alpha} \frac{1}{E_J -E_{\alpha}} \sum_{k_1,k_2}  \sj{F}{J}{I}{k_1}{I}{J_\alpha}\sj{F}{J}{I}{k_2}{I}{J_\alpha} \nonumber \\ 
&& \times (2I+1) \langle I||{\bf T}^{(n)}_{k_1}||I \rangle \langle I||{\bf T}^{(n)}_{k_2}||I \rangle \nonumber \\
&& \times (2J_{\alpha}+1) \langle J_\alpha||{\bf T}^{(e)}_{k_1}||J \rangle \langle J_\alpha||{\bf T}^{(e)}_{k_2}||J \rangle, \label{2ndhyp}
\end{eqnarray}
where we adopt the following definition of the reduced matrix elements \cite{brink1968}:
\begin{eqnarray}
\langle JJ|T_{k}|JJ \rangle = \sqrt{2J+1}\tj{J}{k}{J}{-J}{0}{J} \langle J||{\bf T}_{k}||J \rangle.
\end{eqnarray}
Restricting these expressions to the M1-M1, M1-E2 and E2-E2 interaction terms yields
\begin{align}
E_F^{(2)} &= E_F^{M1-M1} + E_F^{M1-E2} + E_F^{E2-E2} \nonumber \\ 
&= \sum_{\alpha} \left | \sj{F}{J}{I}{1}{I}{J_\alpha} \right |^2 \eta  \nonumber \\
    & \ \qquad + \sj{F}{J}{I}{1}{I}{J_\alpha} \sj{F}{J}{I}{2}{I}{J_\alpha}  \zeta \nonumber \\
    & \ \qquad +  \left | \sj{F}{J}{I}{2}{I}{J_\alpha}   \right |^2 \xi  \label{eq:2ndhyp}
\end{align}
where 
\begin{align}
\eta = & \mu_I^2 \frac{(I+1)(2I+1)}{I} \frac{\Bigl \lvert \sqrt{2J_{\alpha}+1} \langle J_{\alpha} ||{\bf T}^{(e)}_1|| J \rangle \Bigr \rvert^2} {E_{J}- E_{J_{\alpha}}} \notag \\
\xi =& \frac{Q^2}{4}\frac{(I+1)(2I+1)(2I+3)}{I(2I-1)}\frac{\Bigl \lvert \sqrt{2J_{\alpha}+1} \langle J_{\alpha} ||{\bf T}^{(e)}_2|| J \rangle \Bigr \rvert^2} {E_{J}- E_{J_{\alpha}}} \notag \\
\zeta =& \mu_I Q \frac{(I+1)(2I+1)}{I} \sqrt{\frac{2I+3}{2I-1}}  \notag \\ 
       & \frac{(2J_{\alpha}+1)\langle J_{\alpha} ||{\bf T}^{(e)}_1|| J \rangle  \langle J_{\alpha} ||{\bf T}^{(e)}_2|| J \rangle} {E_{J}- E_{J_{\alpha}}}
\end{align}

\subsection{Atomic structure calculations}\label{atomic}

In order to accurately extract the hyperfine $C$-constant and the nuclear octupole moment $\Omega$ from the hyperfine spectra, the electronic matrix element $\langle JJ |T^{e}_3| JJ \rangle$ (see Eq. (\ref{HFScon})) and corrections for second-order effects due to the perturbation of the close-lying $^3$P$^{\circ}_1, ^3$P$^{\circ}_0$ and $^1$P$^{\circ}_1$ states have to be determined. 
In the framework of the multi-configuration Dirac-Hartree-Fock (MCDHF) method~\cite{Grant2007, FroeseFischer2016a}, we calculated the hyperfine interaction constants by using the GRASP2018 package~\cite{FroeseFischer2019} and a new developed RHFS code~\cite{JGLiunpublished} based on  HFS92~\cite{Jonsson1996}. In this work, Yb was treated as a divalent atomic system, that is, $6s$ and $6p$ are the valence orbitals and $1s^22s^22p^63s^23p^63d^{10}4s^24p^64d^{10}5s^25p^64f^{14}$ the core. We started from the Dirac-Hartree-Fock (DHF) calculation, in which all occupied orbitals were optimized in order to minimize the average energy of the of the $6s6p$ configuration minimize. Subsequently, the core-valence (CV) correlations between electrons in the core with $n \ge 4$ and the $6s$ and $6p$ valence electrons were taken into account. To capture the CV correlation effects, the configuration state functions (CSFs), expanding an electronic state $|J M_J \rangle$ concerned, were generated by single (S) and restricted double (D) excitations of electrons from the certain occupied orbitals to a set of virtual orbitals (VOs). The restricted double excitation means that only one electron in the core can be promoted at a time. The VO set was augmented layer by layer, and each layer is composed of orbitals with different angular symmetries, for instance, \textit{s, p, d, \dots}. The VOs were also optimized in the self-consistent field (SCF) procedures, but only those in the last added layer are variable. The CV electron correlation effects on the hyperfine interaction constants were saturated with eight layers of VOs. Keeping all orbitals frozen, we further performed the relativistic configuration interaction (RCI) computations to account for the correlations among the $n=5$ core electrons. It should be stressed that triple (T) and quadrupole (Q) excitations were included in part by the MR-SD approach~\cite{Li2012,Li2016a}. In the MR-SD approach, the CSFs corresponding to TQ excitations were yielded by replacing one and two occupied orbitals in the multi-reference (MR) configurations with the VOs. Here, we selected the $5s^25p^45d^26s6p$ and $5s^25p^66s6p$ configurations to form the MR configuration set. Finally, the Breit interaction was evaluated in RCI as well. In table \ref{atomic_results_1}, we present the calculated $A$, $B/Q$ and $C/\Omega$ as functions of the computational model. 

The ``CC($n=5$)" model stands for the calculation with inclusion of SD-excitation CSFs from the $n=5$ core shell that capture the core-core (CC) electron correlation in this shell. As can be seen, the CV and CC correlations make significant contributions to these constants. Comparing our results with other recent measurement and calculations by Porsev \textit{et al}~\cite{Porsev1999} and Singh \textit{et al.}~\cite{Singh2013}, we found an excellent agreement for the $A$ constant. However, this coincidence may be accidental since $\mu_I = - 0.648(3)$~$\mu_N$ from Ref.~\cite{Stone2005} was adopted in the present calculation, while Mertzimekis \textit{et al.} recommend another nuclear dipole moment, $\mu_I = -0.67989(3)$ for $^{173}$Yb~\cite{Mertzimekis2016}. Multiplied by the ratio of these two nuclear dipole moments, the $A$ constant changes to $-782$~MHz, different from the experimental value by about 5\%. Additionally, the Bohr-Weisskopf effect~\cite{Bohr1950}, which is non-negligible for heavy atoms, was not taken into account. Consequently, we assessed our computational uncertainty to be $\sim 5-10\%$ for the hyperfine interaction constants under investigation. This estimate was further confirmed by comparison of the $B/Q$ constant between ours and Porsev \textit{et al.}. At the same time, it is worth noting the large discrepancies from the calculation by Singh \textit{et al.} for the $B/Q$ and $C/\Omega$ constants~\cite{Singh2013}. In particular, their $C/\Omega$ constant is almost four times larger than our result and, moreover, has the opposite sign. 

Making use of the ``Breit" model, we calculated the off-diagonal hyperfine interaction constants between the $^3$P$^\circ_2$ state and perturbing states $^3$P$^\circ_1$, $^1$P$^\circ_1$, and $^3$P$^\circ_0$. The results are given in Table~\ref{2nd-hfs}.

\begin{table}[!ht]
\caption{Magnetic dipole ($A$ in MHz), electric-field gradient at the nucleus ($B/Q$ in MHz/b) and magnetic octupole ($C/\Omega$ in kHz/($\mu_N \times b)$) hyperfine interaction constants of the $^3$P$^\circ_2$ state in $^{173}$Yb. $\mu_I = -0.648(3)$ $\mu_N$ is taken from Ref.~\cite{Stone2005}. } \label{atomic_results_1}
\begin{ruledtabular}
\begin{tabular}{cccc}
Models             &       A       & $B/Q$ &  $C/\Omega$ \\
\hline
DHF                &     $-579$    &  231  &  2.99   \\
CV($n \ge 4$)      &     $-826$    &  510  &  5.45   \\
CC(n=5)            &     $-767$    &  429  &  4.65   \\
MR(n=5)            &     $-749$    &  434  &  4.65   \\
Breit              &     $-745$    &  432  &  4.51   \\
Porsev \textit{et al.}~\cite{Porsev1999}    &     $-745$             &   477\footnotemark[1]      &                                              \\
Singh \textit{et al.}~\cite{Singh2013}        &     $-742.11$\footnotemark[2]     &   544.6\footnotemark[3]   &     $-15.99$\footnotemark[3]    \\
\end{tabular}
\end{ruledtabular}
\footnotetext[1]{The electric quadrupole moment $Q=2.80$~b was used to extracted the $B/Q$ constant.}
\footnotetext[2]{Experimental values.}
\footnotetext[3]{Theoretical results.}
\end{table}

\begin{table}
\caption{Off-diagonal hyperfine interaction constants and the second-order hfs corrections (in MHz) obtained with the ``Breit" model.  $\mu_I = -0.648(3)$~$\mu_N$ and $Q=2.8$~b are adopted to calculate the second-order hyperfine shifts. The energy intervals involved are taken from the NIST database~\cite{NIST}} \label{2nd-hfs}
\begin{tabular}{r|ccc}
& $^3\!P_1$  &  $^1\!P_1$   & $^3\!P_0$\\
\hline
${\bf T}^{(e)}_1$ (MHz/$\mu_I $) & 4162 &  -7649     &  0 \\
${\bf T}^{(e)}_2$ (MHz/b )       & -670 &  3         &  -815 \\
$\eta$                       & 3.56 & -3.85      &  0 \\
$\zeta$                      & 3.50 & -0.01      &  0 \\
\end{tabular}
\end{table}

\section{Experimental method}\label{experiment}

\subsection{Collinear laser spectroscopy of $^{173}$Yb}

\begin{figure*}[ht]
  \includegraphics[width = 1\textwidth]{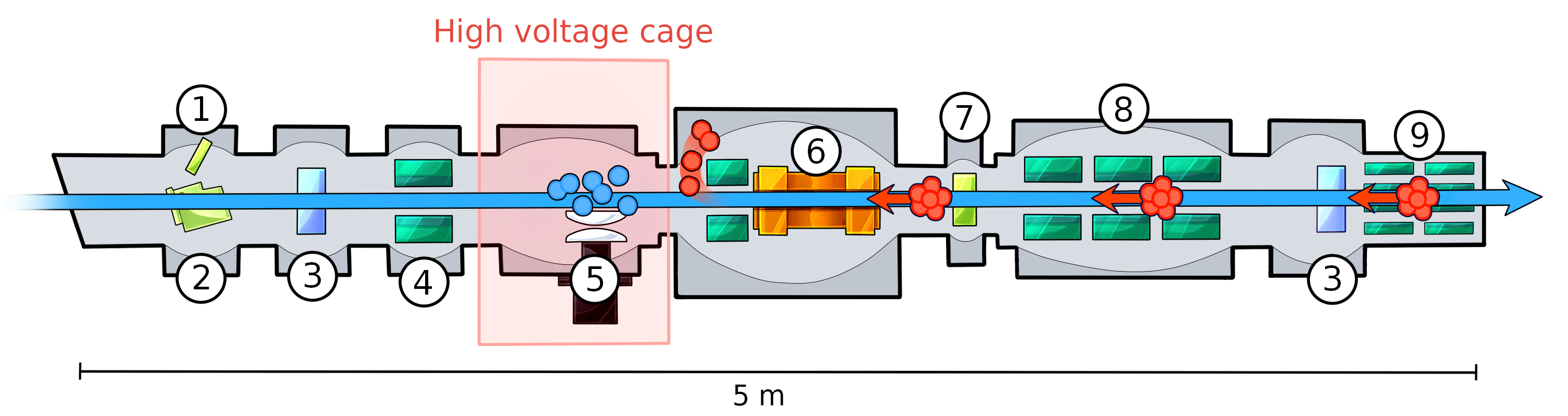}
  \caption{Schematic figure of the collinear laser spectroscopy beamline at IGISOL. The ion beam is injected from the radiofrequency cooler (not shown) into the beamline from the right, while the continuous wave laser overlaps in a counter-propagating direction.  The beamline consists of the following components: 1) Micro-channel plate detector, 2) MagneToF mini detector, 3) Faraday cup, 4) Electrostatic deflector, 5) Photo multiplier tube, 6) Charge-exchange cell,  7) Silicon detector, 8) Quadrupole triplet, 9) XY steerers.}\label{fig:beamline}
\end{figure*}

In order to measure the hyperfine intervals of the metastable $^3$P$^{\circ}_2$ state (located at 19710.388 cm$^{-1}$), the state first needs to be populated. Our method of choice is to rely on a charge-exchange reaction between a fast Yb ion (in our case with a 30\,keV energy) and a neutral K atom. At these higher beam energies, the cross section for this reaction is quite high and sufficiently non-selective to allow several states to be populated. Based on the theoretical calculations presented in \cite{Vernon2019}, a useful fraction of the ions should neutralize into the metastable state of interest, or decay into that state following the few\,$\mu$s travel time between neutralization and detection. Furthermore, because of the kinematic compression of the velocity distribution when accelerating an ion beam from room temperature to 30\,keV, quasi-doppler free spectroscopy is possible by overlapping the ion beam with a laser beam collinearly along the ion beam axis.

The experiment was performed using the collinear laser spectroscopy beamline \cite{vormawah2018} at the IGISOL facility in the Accelerator Laboratory of the University of Jyväskylä, Finland. An intense beam of approximately $10^{10}$Yb ions per second is produced using a glow-discharge spark source. Ions are extracted from this source at an energy of 30\,keV. This beam is guided into a dipole magnet, which is used to mass-select only one of the stable Yb isotopes. The mass resolving power $m/\Delta m$ of this magnet was measured to be 350, which is sufficiently high to ensure the separated beams are isobarically pure. After this mass-selection stage, 
 the ions are injected into a gas-filled radio-frequency Paul trap (rf cooler), where the ions are cooled through collisions with room-temperature helium gas. Ions are extracted from this device as a continuous beam at an energy of 30192(5)\,eV. The method used to determine the absolute value of this beam energy is detailed section C. 
 Drifts and relative changes in the beam energy were monitored using a 1:$10^4$ voltage divider and a digital multimeter, and were found to be well below 1\,V over the duration of the measurement.

The ytterbium ions are transported to the collinear laser beamline as shown in Figure \ref{fig:beamline}, and through a charge-exchange cell, which consists of a reservoir of potassium heated to about 120$^{\circ}$C. Through charge-exchange reactions with the potassium vapour inside this cell, the Yb ions are neutralized. This neutralization process predominantly populates the atomic ground state, but a fraction (estimated at 0.01 percent) of the atoms leaves the charge exchange cell in the $^3$P$^{\circ}_2$ metastable state. Next, the fast atom beam is overlapped with a counter-propagating continuous wave laser beam. The laser light is produced using a Sirah Matisse TS Titanium:Sapphire laser, frequency locked to a commercial HighFinesse WS10 wavemeter. The output of the laser is frequency doubled by a Sirah wavetrain external frequency doubling cavity. By tuning a voltage applied to the charge exchange cell, the velocity of the atoms can be changed. The hyperfine structure of the atoms can thus be measured by doppler-shifting the laser wavelength observed in the rest frame of the atoms. The acceleration voltage is measured using a 1:$10^3$ voltage divider and a digital multimeter for calibration, which prevents e.g. non-linearities in the voltage scan.

\begin{figure*}[ht]
  \includegraphics[width = 1\textwidth]{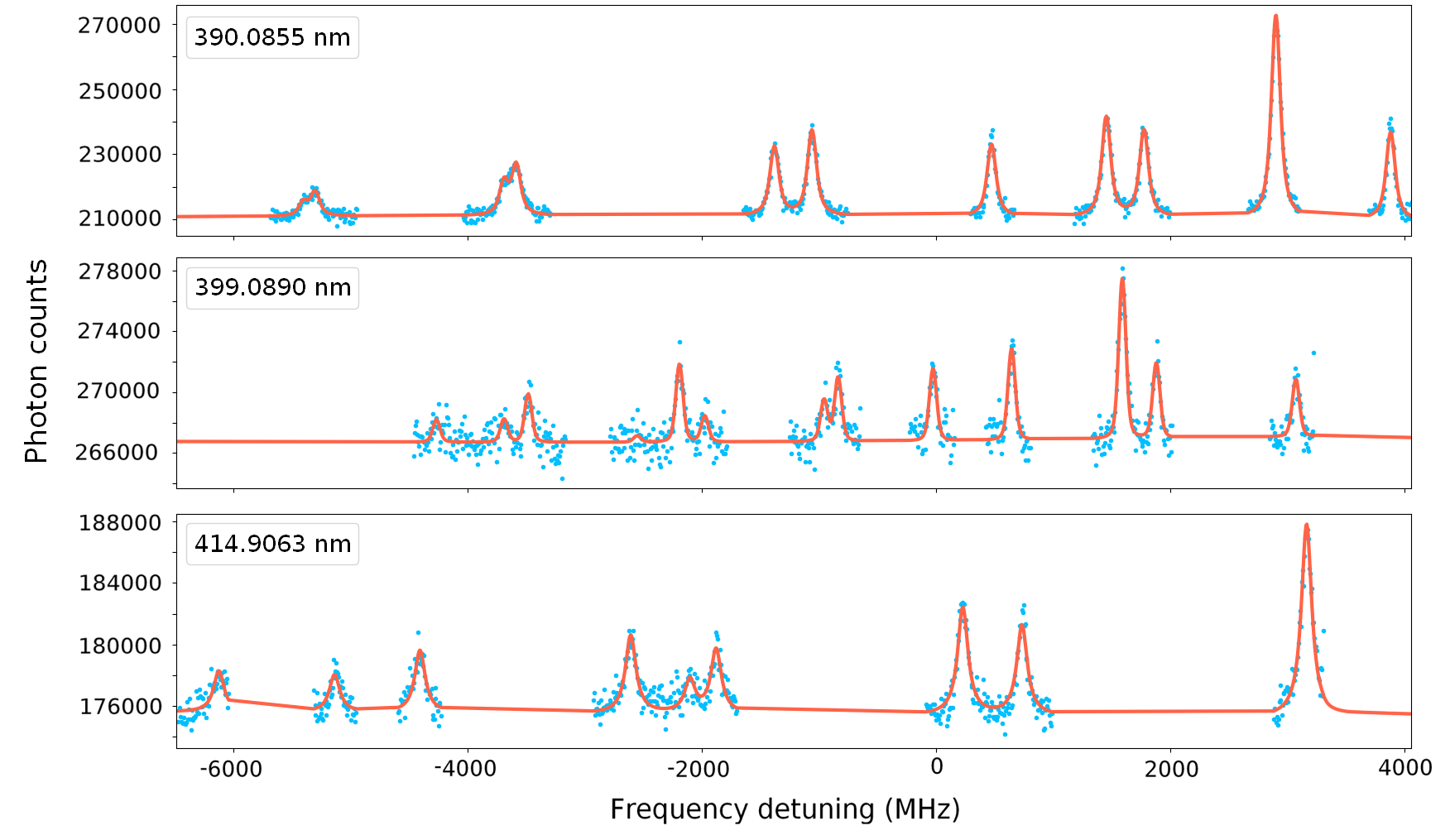}
  \caption{Typical spectra obtained with the three optical transitions from the metastable state of interest used in this work. 
  The red line is the line of best fit.}\label{fig:spectra}
\end{figure*}

Spectroscopy was performed using three different optical transitions from the $^3$P$^{\circ}_2$ state, as well as using one transition from the 4f$^{14}$6s$^2$ $^1$S$_0$ ground state.  These lines were the 399.0890\,nm, 390.0855\,nm and 414.9063\,nm transitions (wavelengths in vacuum) to the 4f$^{14}$6p$^2$ $^3$P$_2$, 4f$^{13}(^2$F$^\circ_{7/2}$) 5d($^2$D)6s6p($^3$P$^\circ$) ($^2$D$^{\circ}_{5/2}$) ($J=2$) and 4f$^{14}$6p$^2$ $^3$P$_1$ states at 44760.370\,cm$^{-1}$, 45338.530\,cm$^{-1}$ and 43805.42\,cm$^{-1}$. The optical transition from the ground state is at 398.7986\,nm, to the 4f$^{14}$6s6p $^3$D$_1$ state at 25068.222\,cm$^{-1}$. After data processing, discussed in the next section, spectra similar to those shown in Fig.~\ref{fig:spectra} are obtained for the different atomic lines. From our measurements we estimate the ratio of atoms in the $^3$P$^{\circ}_2$ state to atoms in the 4f$^{14}$6s$^2$ $^1$S$_0$ ground state is about 1:$10^4$, in significant disagreement with the theoretical estimate~\cite{Vernon2019}. Despite this low number of metastable atoms, clear hyperfine structures can nevertheless be recorded in a few hours.


\subsection{Analysis procedure}

The measurement procedure outlined above yields the number of photon counts observed by the photomultipler tube (PMT) for every acceleration voltage in the sweep. Using this calibrated voltage, the laser frequency as observed in the rest frame of the atoms can be calculated via the relativistic Doppler shift formula:
\begin{align}
    \nu_{\text{rest}} &= \nu_{\text{lab}} \frac{1+\beta}{\sqrt{1-\beta^2}},\\
    \beta &= \frac{v}{c} = \sqrt{1 - \frac{m^2c^4}{(eV + mc^2)^2}}, \notag
\end{align}
where $V$ is the sum of the cooler platform voltage and the acceleration voltage $V = V_{\text{cool}} + V_{\text{acc}}$, and $\nu_{\text{lab}}$ is the laser frequency as measured by the wavemeter. Hyperfine constants can then extracted from the obtained hyperfine spectra. For this, the SATLAS package~\cite{gins2018} was used. The spectra are fit by using Voigt profiles centered at the resonance locations predicted from Eqs.~\eqref{eq:hyp}, corrected for second-order shifts given by Eq.\eqref{eq:2ndhyp}.

Optimal values of the hyperfine constants, linewidths, peak heights and background are then obtained by performing a $\chi^2$-minimization. Uncertainties are estimated from the inverse of the Hessian matrix, and are scaled with $\sqrt{\chi^2_{\text{red}}}$, where $\chi^2_{\text{red}}$ is obtained by dividing $\chi^2$ with the number of degrees of freedom. Each fit has seven free parameters related to the hyperfine structure constants ($A$, $B$ and $C$ for each state, and furthermore a centroid wavelength). Since there are more than seven resonances for each of the transitions, the fit is over-determined, which helps to reduce the statistical uncertainties and to get precise values of the hyperfine coefficients, despite the somewhat poor signal-to-background for some scans. 

The photon background was observed to vary over the scan range. The background is dominated by light spontaneously emitted by the ytterbium atoms following charge exchange. Scanning over large voltage ranges causes small changes in the atom trajectories, which changes the number of photons detected by the PMT. In order to fit this background, a second-order polynomial was used. Note that this choice of background shape affected the values of the hyperfine constants imperceptibly.
For some of the spectra, partially overlapping resonances can be observed. The ratio of amplitudes of such overlapping peaks can be constrained to the ratio of the following theoretical intensities~\cite{Campbell2016}:
\begin{equation}
    I_{FF'} \propto (2F+1)(2F'+1) { \begin{Bmatrix}
   F & F' & 1 \\
   J & J' & I 
  \end{Bmatrix}}^2.
\end{equation}
This helps to ensure convergence of the fit, and prevents convergence to nonphysically small amplitudes for one of the components.

\subsection{Systematic Uncertainties}\label{sec:syst}

\begin{table*}[ht!]
    \centering
    \begin{tabular}{c|ccc|ccc}
        \hline
        \hline
        & A$_{\text{l}}$ [MHz] & B$_{\text{l}}$ [MHz] & C$_{\text{l}}$ [MHz]  & A$_{\text{u}}$ [MHz] & B$_{\text{u}}$ [MHz] & C$_{\text{u}}$ [MHz]\\
        \hline
        \hline
        399.0890\,nm  &  -742.0(1) & 1347.2(10) & -0.01(7)  & -123.4(10) & -629.8(9)   & 0.08(7)\\
        390.0855\,nm&  -741.8(2) & 1348.9(19) & 0.12(13) & -347.6(2) & 120.0(18) & -0.15(14) \\
        414.9063\,nm  &  -742.0(3) & 1345.4(23) & 0.01(17) & -0.01(45)   & 484.0(12)    & -- \\
        \hline
         All lines & -741.98(9) & 1347.3(8) & 0.02(6) & -- & -- & --\\
        \hline
        \hline
        399.0890\,nm (corr) & -742.0(1) & 1347.2(11) & -0.02(9)  & -123.5(12) & -629.9(11)   & 0.09(7)\\
        390.0855\,nm (corr) & -741.8(2) & 1348.9(19) & 0.16(14) & -347.6(2) & 119.9(18) & -0.15(14) \\
        414.9063\,nm (corr) & -742.0(3) & 1345.4(23) & 0.01(17) & -0.01(45)   & 484.0(12)    & -- \\
        \hline
        All lines (corr) & -741.98(9) & 1347.3(9) & 0.03(7) & -- & -- & --\\
        \hline
        \hline
        769.9487\,nm \cite{Singh2013} & -742.11(2) & 1339.2(2) & 0.54(2) & -- & -- & -- \\
    \end{tabular}
    \caption{Summary of the hyperfine constants extracted in this work. Uncertainties (one standard deviation) are given in brackets, and are calculated as the uncertainty on the weighted mean of the values obtained for every separate scan. Values with and without corrections for second-order hyperfine interactions are given.}
    \label{tab:hyperfine}
\end{table*}

One possible source of systematic uncertainty stems from the precision with which the ion beam energy can be determined.. Assuming an incorrect beam energy during the analysis will lead to a small stretching of the frequency scale of the hyperfine structure in the rest frame of the atoms, which in turn may skew the extracted hyperfine intervals. This is not expected to strongly influence the value of the hyperfine $C$-constant, since the presence of a non-zero $C$ only manifests itself as a minor perturbation on top of the much larger dipole and quadrupole splittings. Nevertheless, care was taken to remove this potential systematic effect. This was done via a beam energy calibration using reference measurements of the accurately known hyperfine intervals of the ground state of singly-ionized $^{171,173}$Yb$^+$ \cite{Degroote2019}. 
In order to study the ionic ground state, after collecting several scans for each of the atomic transitions, the charge-exhange cell was brought to room temperature. The ion beam thus was no longer neutralized, and collinear laser spectroscopy was performed on the ionic 369.4192\,nm line from 4f$^{14}$6s $^2$S$_{1/2}$ to the 4f$^{14}$6p $^2$P$^{\circ}_{1/2}$ state at 27061.82\,cm$^{-1}$. Three scans were taken for both $^{171,173}$Yb$^+$. In the case of the present measurements, if the assumed beam energy is wrong by 10\,eV, the extracted hyperfine intervals would be offset from literature values by about 2\,MHz. By combining the three scans for $^{171}$Yb and $^{173}$Yb, we obtain a voltage offset of 34(5)\,V. This value was added to the total beam energy during the analysis. The systematic uncertainty on the hyperfine constants due to the 5\,V uncertainty on the beam energy was also estimated and found to be negligible. 

Care was also taken to optimize the mass resolving power of the dipole magnet, to make sure no $^{172,174}$Yb is present in the beam when studying $^{173}$Yb. Otherwise, if not accounted for, e.g. a small resonance of the even-even neighbours which lies close to a $^{173}$Yb resonance might affect the hyperfine constants. No evidence for resonances associated with these neighbouring masses (whose position is well known from separate measurements performed prior to measuring the structure of $^{173}$Yb) was seen in the spectra. 

As will be shown in section\ref{conclusion}, the data extracted for the three different transitions all yield consistent values. This provides a measure of confidence in the final hyperfine constants extracted here. It should be noted however that any source of systematic uncertainty which affects all three lines equally would remain undetected.

\section{Results and discussion}\label{conclusion}

\subsection{Hyperfine constants of the $^3$P$^{\circ}_2$-state}

\begin{figure*}[ht]
  \includegraphics[width = 1\textwidth]{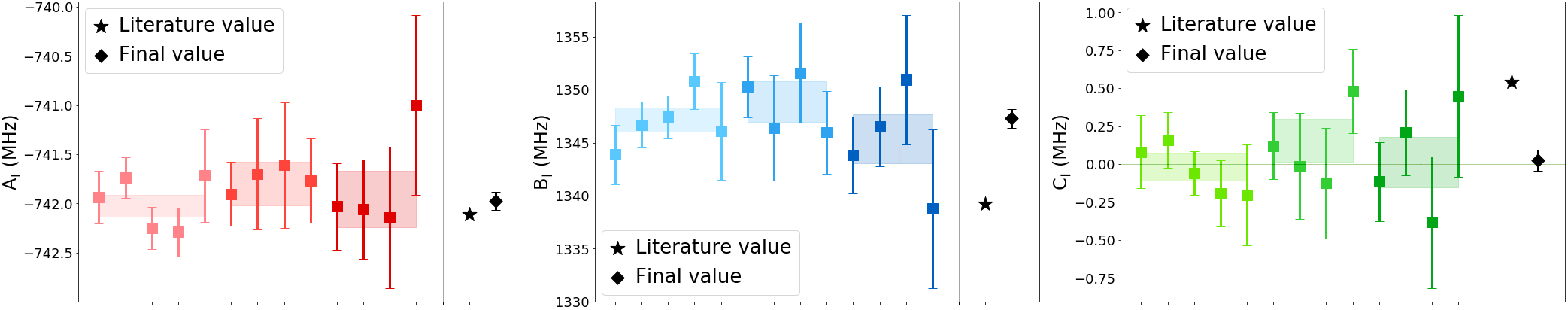}
  \caption{Summary of the hyperfine $A,B,$ and $C$-constants extracted from the 13 measurements performed in this work. The weighted mean is shown with the shaded areas. For the $C$-constants, our final value is zero within the uncertainties, and significantly different from the literature value.}\label{fig:hypconstants}
\end{figure*}

The hyperfine constants extracted from the measurements are summarized in Table~\ref{tab:hyperfine}. The numbers in round brackets are statistical and are calculated as the error on the weighted mean of the results for the individual scans of each transition. The fourth row and eighth row of the table show the weighted average of the three preceding rows. Values are presented with and without taking the second-order shifts discussed in section \ref{atomic} into account. The impact of the second-order shifts is rather small given our experimental uncertainties - the value of the hyperfine $C$-constant shifts from 0.02(6)\,MHz to 0.03(7)\,MHz, but the average values of $A$ and $B$ do not change at the present precision. The hyperfine constants are also compared to literature. Good agreement with literature is obtained for the hyperfine $A$ coefficient, but the value of both $B$ and $C$ deviate. Interestingly, the hyperfine $A$-constant of the 4f$^{14}$6p$^2$ $^3$P$_1$ state is zero within our uncertainties.

The hyperfine $C$-constant of the $^3$P$^{\circ}_2$ state is zero within one standard deviation for every individual line. The weighted mean of the values of $C$ for this state from all 13 scans is 0.03(7)\,MHz (at the $68\%$ confidence level), which is significantly different from the value of 0.54(2)\,MHz obtained using the 769.9487\,nm line \cite{Singh2013}. This is shown graphically in the third panel of Fig.~\ref{fig:hypconstants}, which shows the values of $A, B$ and $C$ for every individual scan, as well as their weighted average. Fig.~\ref{fig:hypconstants} serves to illustrate the excellent internal consistency of our dataset. Given that three different optical transitions were used to study the hyperfine structure of the $^3$P$^{\circ}_2$ state, this provides confidence in our values of the hyperfine constants. The origins of the discrepancy between our value of $C$ and literature are unclear; further investigation using the 769.9487\,nm transition in the future might help to resolve this puzzle. Currently, the wavelength of all light spontaneously emitted by the atom following the 769.9487\,nm excitation lies outside of the wavelength sensitivity of our photo-multiplier tube, preventing us from making this measurement at the present time. 

In order to extract a value of $\Omega$ from our new value of $C$ for $^{173}$Yb, atomic structure calculations are required. The value of $C/\Omega$ was reported as $-15.99$\,kHz/$(\mu_N \times b)$ in~\cite{Singh2013}. Using this value, we thus obtain -1.9(44)\,$\mu_N \times b$. From the calculations presented in this work (see Table~\ref{atomic_results_1}), we find $C/\Omega = 4.51$\,kHz/($\mu_N \times$ b), yielding 6.7(160)\,$\mu_N \times b$. 

\subsection{Discussion}

\begin{figure}
  \includegraphics[width = \columnwidth]{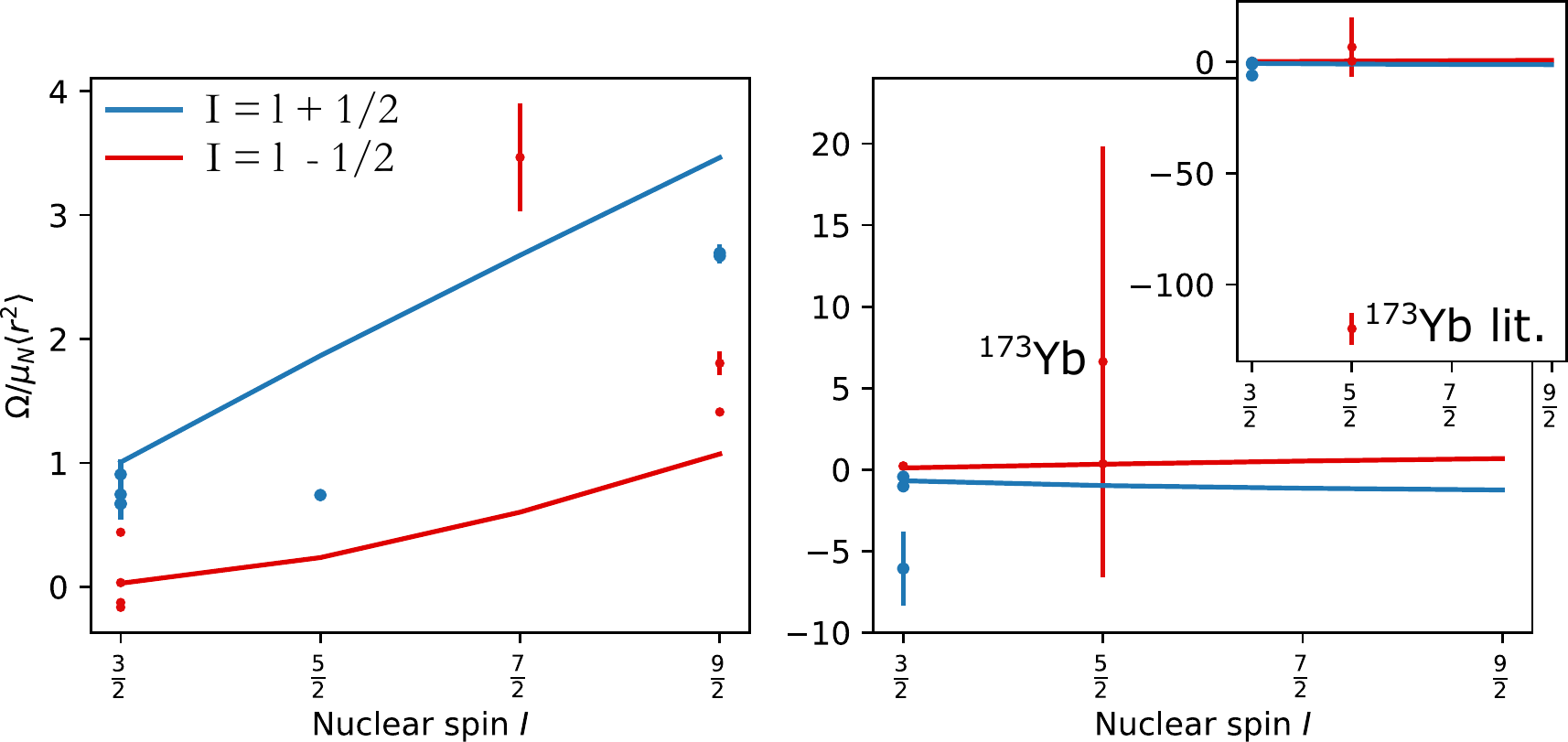}
  \caption{Overview of the currently known magnetic octupole moments of odd-mass isotopes (listed in table~\ref{tab:octupole}). Full lines are single-particle estimates of $\Omega$ obtained using equation~\eqref{eq:single_particle_moment}, with an orbital quenching factor of 0.6. Left: odd-proton isotopes, right: odd-neutron nuclei. The inset shows a zoomed-out scale, to bring the literature value for $^{173}$Yb into view.}\label{fig:overview}
\end{figure}

The value of $\Omega = 34.4(21)\,\mu_N$b obtained in~\cite{Singh2013} was found to be many times larger than the simple shell model, which formed the motivation for the current work. An overview of the experimental values of $\Omega$ available in the literature is given in Tab.~\ref{tab:octupole}, which serves to place into context the size of the value of $\Omega$ of $^{173}$Yb obtained in~\cite{Singh2013}. An overview of the existing experimental values of $\Omega/\mu_N \left\langle r^2 \right\rangle$, where $\left\langle r^2 \right\rangle$ is the mean-squared radius of the valence nucleon orbital, is furthermore shown in Fig.~\ref{fig:overview}. Error bars represent the total uncertainties as reported in literature. To obtain values of $\left\langle r^2 \right\rangle$, in principle theoretical calculations would be required, but as an approximation data on the total nuclear charge radius obtained from~\cite{Angeli2013} were used. Fig.~\ref{fig:overview} also shows theoretical single-particle estimates of $\Omega/\mu_N \left\langle r^2 \right\rangle$ of a nucleus with spin $I$ containing an odd number of nucleons. These values can be calculated using~\cite{Schwartz1955}
\begin{align}
    \Omega/\mu_N \left\langle r^2 \right\rangle &=  \frac32 \frac{2I-1}{(2I+4)(2I+2)} \notag\\
             & \times \begin{cases}
                          (I+2)[(I-\frac32)g_l + g_s], & I=l+\frac12 \\
                          (I-1)[(I+\frac52)g_l - g_s], & I=l-\frac12
                       \end{cases} \label{eq:single_particle_moment}
\end{align}
where $g_l=1, g_s= 3.351414$ for free protons and , $g_l=0, g_s= -2.2961127$ for free neutrons. From the systematics of the magnetic dipole moments, it is well known that these free nucleon values do not provide an adequate description for the vast majority of nuclei. For this reason, often quenched values, usually $g_{s,\text{eff}} = 0.6\cdot g_{s,\text{free}}$, are adopted. This is also the approach we take here for the magnetic octupole moment. 

Most values of $\Omega/\mu_N \left\langle r^2 \right\rangle$ lie between the curves predicted using Eq.\,\eqref{eq:single_particle_moment}, similar to the case for the magnetic dipole moments as well, with a few exceptions. As noted in \cite{Gerginov2003}, the value for $^{133}$Cs is markedly larger than both the single-particle estimate without quenching, and even the quenched single-particle estimate is considerably smaller than the experimental value. Another example is reported for $^{155}$Gd \cite{Unsworth1969}, where  $\Omega$ is significantly larger than the quenched single-particle estimate (0.029\,$\mu_N b$) and the predictions by the strong coupling model (0.044\,$\mu_N b$) \cite{Williams1962}. However, the Gd atom has an opened $d$ and half-filled $f$ subshells. Thus, it poses a significant challenge for atomic-structure calculations to deal with the extremely complicated electron correlations and to achieve the required accuracy for extraction of $\Omega$ from hyperfine splitting measurements. 

From Fig.~\ref{fig:overview}, it is clear that our new value of $\Omega = 6.7(16)\,\mu_N$ b is consistent with the simple shell model estimate. More precise measurements of $\Omega$ are required in order to make more detailed comparisons with the shell model picture presented here, or with future large-scale nuclear structure calculations. For example, advanced large-scale shell model calculations are nowadays routinely used to accurately investigate the finer details of the magnetic dipole moments. For magnetic octupole moments such calculations would also be desirable, though so far only few attempts have been made~\cite{brown1980}. 

\begin{table}[!ht]
\centering
\setlength{\tabcolsep}{3pt}
\caption{Overview of some of the magnetic octupole hyperfine constants $C$ and their corresponding moments. Error bars on the values of $C$ and $\Omega$ are as quoted in the literature, where available. $\mu_N$ is the nuclear magneton.}\label{tab:octupole}
\begin{threeparttable}
\begin{tabular}{c|cccccc}
    \toprule
    Isotope	& Z	& N & State & $C$ [Hz] & $\Omega [\mu_N b]$  & Ref. \\[3pt]
    \hline
    \hline
    $^{35}$Cl  & 17 & 18  &	$^2$P$^{\circ}_{3/2}$ & -6.95(1.2) & -0.0188    & \tnote{a} \\
    $^{37}$Cl  & 17 & 20  &	$^2$P$^{\circ}_{3/2}$ & -5.41(1.2) & -0.0146    & \tnote{a} \\
    $^{51}$V   & 23 & 27  & $^4$D$_{3/2} $        & -1660(140) &    --      & \cite{Childs1978} \tnote{b} \\
               & 23 & 27  & $^4$D$_{5/2} $        & -270(80)   &    --      & \cite{Childs1978} \tnote{b} \\
    $^{69}$Ga  & 31 & 38  &	$^2$P$^{\circ}_{3/2}$ & 84(6)	   & 0.107(20)	& \cite{Daly1954} \\
    $^{71}$Ga  & 31 & 40  &	$^2$P$^{\circ}_{3/2}$ & 115(7)     & 0.146(20) 	& \cite{Daly1954} \\
    $^{79}$Br  & 35 & 44  &	$^2$P$^{\circ}_{3/2}$ & 388(8)	   & 0.116		& \cite{Brown1966} \\
    $^{81}$Br  & 35 & 46  &	$^2$P$^{\circ}_{3/2}$ & 430(8)	   & 0.129		& \cite{Brown1966} \\
    $^{83}$Kr  & 36 & 47  &	$^3$P$^{\circ}_2$     & -790(250)   & -0.18(6)	& \cite{Faust1963} \\
    $^{113}$In & 49 & 64  &	$^2$P$^{\circ}_{3/2}$ & 1728(45)   & 0.565(12)	& \cite{Kusch1957} \\
    $^{115}$In & 49 & 66  &	$^2$P$^{\circ}_{3/2}$ & 1702(35)   & 0.574(15)	& \cite{Kusch1957} \\
    $^{127}$I  & 53 & 74  & $^2$P$^{\circ}_{3/2}$ & 2450(370)  & 0.167      & \cite{Jaccarino1954,Amoruso1971} \\
    $^{131}$Xe & 54 & 77  &	$^3$P$^{\circ}_2$     & 728(105)   & 0.048(12)	& \cite{Faust1961} \\
    $^{133}$Cs & 55 & 78  &	$^2$P$^{\circ}_{3/2}$ & 560(70)	   & 0.8(1)	    & \cite{Gerginov2003} \\
    $^{137}$Ba & 56 & 81  &	$^{2}$D$_{3/2}$       & 36.546     & 0.05057(54)& \cite{Lewty2012} \\
    $^{137}$Ba & 56 & 81  &	$^{2}$D$_{5/2}$       & -12.41(77) & 0.0496(37)& \cite{Lewty2013} \\
    $^{151}$Eu & 63 & 88  & $^{8,10}$D$^{\circ}_J$& \tnote{c}  & 	--   	& \cite{Childs1991} \\
    $^{153}$Eu & 63 & 90  & $^{8,10}$D$^{\circ}_J$& \tnote{c}  & 	--  	& \cite{Childs1991} \\
    $^{155}$Gd & 64 & 91  & $^3$D$^{\circ}_3$	  & -1500(500) & -1.6(6)	& \cite{Unsworth1969} \\
    $^{173}$Yb & 70 & 103 & $^3$P$^{\circ}_2$	  & 540(20)$\cdot 10^3$	& -34.4(21)	& \cite{Singh2013} \\
    $^{175}$Lu & 71 & 104 &	                      & -617(57)   & 	--  	& \cite{Brenner1985} \\
    $^{176}$Lu & 71 & 105 & 	                  & -1377(302) & 	--  	& \cite{Brenner1985} \\
    $^{177}$Hf & 72 & 105 & $^3$F$_2$	          & 170(30)	   & 	--   	& \cite{Jin1995} \\
               & 72 & 105 & $^3$F$_3$	          & 230(90)	   & 	--   	& \cite{Jin1995} \\
               & 72 & 105 & $^3$F$_4$	          & 500(190)   & 	--   	& \cite{Jin1995} \\
    $^{179}$Hf & 72 & 107 & $^3$F$_2$             & -410(50)   & 	--   	& \cite{Jin1995} \\
               & 72 & 107 & $^3$F$_3$             & -600(250)  & 	--   	& \cite{Jin1995} \\
               & 72 & 107 & $^3$F$_4$             & -1190(500) & 	--   	& \cite{Jin1995} \\
    $^{197}$Au & 79 & 118 & $^2$D$_{3/2}$	      & 212(14)	   & 0.0098(7)	& \cite{Blachman1967} \\
       & 79 & 118 & $^4$F$^{\circ}_{9/2}$	      & 326(10)    & 0.130		& \cite{Blachman1967} \\
    $^{201}$Hg & 80 & 121 & $^3$P$^{\circ}_2$     & -1840(90)  & -0.130(13)	& \cite{McDermott1960} \\
    $^{209}$Bi & 83 & 126 & $^2$P$^{\circ}_{3/2}$ & 19300(500) & 0.55(3)\tnote{d} & \cite{Landman1970}\\
               & 83 & 126 & $^4$S$^{\circ}_{3/2}$ & 16500(100) & 0.43       & \cite{hull1970}\\
    $^{207}$Po & 84 & 123 & $^3$P$_2$             & -12(1)	   & 0.11(1)    & \cite{olsmats1961,fuller1976}\\
\end{tabular}
\begin{tablenotes}
    \item[a] J. H. Holloway, unpublished Ph.D. thesis, Department of Physics, MIT, 1956. Values taken from \cite{Brown1966}.
    \item[b] Listing only significantly non-zero $C$-values.
    \item[c] Average ratio of $C$-constants provided: C($^{151}$Eu)/C($^{153}$Eu) = 0.87(6).
    \item[d] Value and uncertainty obtained as the average of several theoretical methods.
\end{tablenotes}
\end{threeparttable}
\end{table}

\section{Conclusion}

We reported on the measurement of the hyperfine constants $A,B, C$ of the $^3$P$^{\circ}_2$ state in neutral Yb. These measurements were performed using three different transitions with a standard collinear laser spectroscopy method, where the state of interest was populated through charge exchange with neutral potassium. From the data, we find that the hyperfine $C$-constant is zero within experimental uncertainties. Our measurement thus resolves a puzzlingly large value reported in literature, which seemed difficult to reconcile with simple shell model estimates of the magnetic octupole moment. Future work to obtain the value of this magnetic octupole moment will require at least an order of magnitude improvement in precision. This could be achieved using e.g. a collinear laser-rf method, or another suitably high-precision technique. We also reported on a set of atomic structure caluclations of the second-order hyperfine shifts and the value of $C/\Omega$ which will be required to interpret such future, higher-precision measurements.


\begin{acknowledgments}

R.P.D.G. received funding from the European Union's Horizon 2020 research and innovation programme under the Marie Sk{\l}odowska-Curie grant agreement No 844829. J.G.L. acknowledges the support of the National Natural Science Foundation of China under Grant No. 11874090.

\end{acknowledgments}


\begin{thebibliography}{46}%
\makeatletter
\providecommand \@ifxundefined [1]{%
 \@ifx{#1\undefined}
}%
\providecommand \@ifnum [1]{%
 \ifnum #1\expandafter \@firstoftwo
 \else \expandafter \@secondoftwo
 \fi
}%
\providecommand \@ifx [1]{%
 \ifx #1\expandafter \@firstoftwo
 \else \expandafter \@secondoftwo
 \fi
}%
\providecommand \natexlab [1]{#1}%
\providecommand \enquote  [1]{``#1''}%
\providecommand \bibnamefont  [1]{#1}%
\providecommand \bibfnamefont [1]{#1}%
\providecommand \citenamefont [1]{#1}%
\providecommand \href@noop [0]{\@secondoftwo}%
\providecommand \href [0]{\begingroup \@sanitize@url \@href}%
\providecommand \@href[1]{\@@startlink{#1}\@@href}%
\providecommand \@@href[1]{\endgroup#1\@@endlink}%
\providecommand \@sanitize@url [0]{\catcode `\\12\catcode `\$12\catcode
  `\&12\catcode `\#12\catcode `\^12\catcode `\_12\catcode `\%12\relax}%
\providecommand \@@startlink[1]{}%
\providecommand \@@endlink[0]{}%
\providecommand \url  [0]{\begingroup\@sanitize@url \@url }%
\providecommand \@url [1]{\endgroup\@href {#1}{\urlprefix }}%
\providecommand \urlprefix  [0]{URL }%
\providecommand \Eprint [0]{\href }%
\providecommand \doibase [0]{http://dx.doi.org/}%
\providecommand \selectlanguage [0]{\@gobble}%
\providecommand \bibinfo  [0]{\@secondoftwo}%
\providecommand \bibfield  [0]{\@secondoftwo}%
\providecommand \translation [1]{[#1]}%
\providecommand \BibitemOpen [0]{}%
\providecommand \bibitemStop [0]{}%
\providecommand \bibitemNoStop [0]{.\EOS\space}%
\providecommand \EOS [0]{\spacefactor3000\relax}%
\providecommand \BibitemShut  [1]{\csname bibitem#1\endcsname}%
\let\auto@bib@innerbib\@empty
\bibitem [{\citenamefont {Neyens}(2003)}]{neyens2003}%
  \BibitemOpen
  \bibfield  {author} {\bibinfo {author} {\bibfnamefont {G.}~\bibnamefont
  {Neyens}},\ }\href@noop {} {\bibfield  {journal} {\bibinfo  {journal}
  {Reports on progress in physics}\ }\textbf {\bibinfo {volume} {66}},\
  \bibinfo {pages} {633} (\bibinfo {year} {2003})}\BibitemShut {NoStop}%
\bibitem [{\citenamefont {Campbell}\ \emph {et~al.}(2016)\citenamefont
  {Campbell}, \citenamefont {Moore},\ and\ \citenamefont
  {Pearson}}]{Campbell2016}%
  \BibitemOpen
  \bibfield  {author} {\bibinfo {author} {\bibfnamefont {P.}~\bibnamefont
  {Campbell}}, \bibinfo {author} {\bibfnamefont {I.}~\bibnamefont {Moore}}, \
  and\ \bibinfo {author} {\bibfnamefont {M.}~\bibnamefont {Pearson}},\ }\href
  {\doibase https://doi.org/10.1016/j.ppnp.2015.09.003} {\bibfield  {journal}
  {\bibinfo  {journal} {Progress in Particle and Nuclear Physics}\ }\textbf
  {\bibinfo {volume} {86}},\ \bibinfo {pages} {127 } (\bibinfo {year}
  {2016})}\BibitemShut {NoStop}%
\bibitem [{\citenamefont {Mertzimekis}\ \emph {et~al.}(2016)\citenamefont
  {Mertzimekis}, \citenamefont {Stamou},\ and\ \citenamefont
  {Psaltis}}]{Mertzimekis2016}%
  \BibitemOpen
  \bibfield  {author} {\bibinfo {author} {\bibfnamefont {T.}~\bibnamefont
  {Mertzimekis}}, \bibinfo {author} {\bibfnamefont {K.}~\bibnamefont {Stamou}},
  \ and\ \bibinfo {author} {\bibfnamefont {A.}~\bibnamefont {Psaltis}},\ }\href
  {\doibase http://dx.doi.org/10.1016/j.nima.2015.10.096} {\bibfield  {journal}
  {\bibinfo  {journal} {Nuclear Instruments and Methods in Physics Research
  Section A: Accelerators, Spectrometers, Detectors and Associated Equipment}\
  }\textbf {\bibinfo {volume} {807}},\ \bibinfo {pages} {56} (\bibinfo {year}
  {2016})}\BibitemShut {NoStop}%
\bibitem [{\citenamefont {Daly}\ and\ \citenamefont
  {Holloway}(1954)}]{Daly1954}%
  \BibitemOpen
  \bibfield  {author} {\bibinfo {author} {\bibfnamefont {R.~T.}\ \bibnamefont
  {Daly}}\ and\ \bibinfo {author} {\bibfnamefont {J.~H.}\ \bibnamefont
  {Holloway}},\ }\href {\doibase 10.1103/PhysRev.96.539} {\bibfield  {journal}
  {\bibinfo  {journal} {Phys. Rev.}\ }\textbf {\bibinfo {volume} {96}},\
  \bibinfo {pages} {539} (\bibinfo {year} {1954})}\BibitemShut {NoStop}%
\bibitem [{\citenamefont {Jaccarino}\ \emph {et~al.}(1954)\citenamefont
  {Jaccarino}, \citenamefont {King}, \citenamefont {Satten},\ and\
  \citenamefont {Stroke}}]{Jaccarino1954}%
  \BibitemOpen
  \bibfield  {author} {\bibinfo {author} {\bibfnamefont {V.}~\bibnamefont
  {Jaccarino}}, \bibinfo {author} {\bibfnamefont {J.~G.}\ \bibnamefont {King}},
  \bibinfo {author} {\bibfnamefont {R.~A.}\ \bibnamefont {Satten}}, \ and\
  \bibinfo {author} {\bibfnamefont {H.~H.}\ \bibnamefont {Stroke}},\ }\href
  {\doibase 10.1103/PhysRev.94.1798} {\bibfield  {journal} {\bibinfo  {journal}
  {Phys. Rev.}\ }\textbf {\bibinfo {volume} {94}},\ \bibinfo {pages} {1798}
  (\bibinfo {year} {1954})}\BibitemShut {NoStop}%
\bibitem [{\citenamefont {Eck}\ and\ \citenamefont {Kusch}(1957)}]{Kusch1957}%
  \BibitemOpen
  \bibfield  {author} {\bibinfo {author} {\bibfnamefont {T.~G.}\ \bibnamefont
  {Eck}}\ and\ \bibinfo {author} {\bibfnamefont {P.}~\bibnamefont {Kusch}},\
  }\href {\doibase 10.1103/PhysRev.106.958} {\bibfield  {journal} {\bibinfo
  {journal} {Phys. Rev.}\ }\textbf {\bibinfo {volume} {106}},\ \bibinfo {pages}
  {958} (\bibinfo {year} {1957})}\BibitemShut {NoStop}%
\bibitem [{\citenamefont {McDermott}\ and\ \citenamefont
  {Lichten}(1960)}]{McDermott1960}%
  \BibitemOpen
  \bibfield  {author} {\bibinfo {author} {\bibfnamefont {M.~N.}\ \bibnamefont
  {McDermott}}\ and\ \bibinfo {author} {\bibfnamefont {W.~L.}\ \bibnamefont
  {Lichten}},\ }\href {\doibase 10.1103/PhysRev.119.134} {\bibfield  {journal}
  {\bibinfo  {journal} {Phys. Rev.}\ }\textbf {\bibinfo {volume} {119}},\
  \bibinfo {pages} {134} (\bibinfo {year} {1960})}\BibitemShut {NoStop}%
\bibitem [{\citenamefont {Olsmats}\ \emph {et~al.}(1961)\citenamefont
  {Olsmats}, \citenamefont {Axensten},\ and\ \citenamefont
  {Liljegren}}]{olsmats1961}%
  \BibitemOpen
  \bibfield  {author} {\bibinfo {author} {\bibfnamefont {C.}~\bibnamefont
  {Olsmats}}, \bibinfo {author} {\bibfnamefont {S.}~\bibnamefont {Axensten}}, \
  and\ \bibinfo {author} {\bibfnamefont {G.}~\bibnamefont {Liljegren}},\
  }\href@noop {} {\bibfield  {journal} {\bibinfo  {journal} {Arkiv Fysik}\
  }\textbf {\bibinfo {volume} {19}} (\bibinfo {year} {1961})}\BibitemShut
  {NoStop}%
\bibitem [{\citenamefont {Faust}\ and\ \citenamefont
  {McDermott}(1961)}]{Faust1961}%
  \BibitemOpen
  \bibfield  {author} {\bibinfo {author} {\bibfnamefont {W.~L.}\ \bibnamefont
  {Faust}}\ and\ \bibinfo {author} {\bibfnamefont {M.~N.}\ \bibnamefont
  {McDermott}},\ }\href {\doibase 10.1103/PhysRev.123.198} {\bibfield
  {journal} {\bibinfo  {journal} {Phys. Rev.}\ }\textbf {\bibinfo {volume}
  {123}},\ \bibinfo {pages} {198} (\bibinfo {year} {1961})}\BibitemShut
  {NoStop}%
\bibitem [{\citenamefont {Faust}\ and\ \citenamefont
  {Chow~Chiu}(1963)}]{Faust1963}%
  \BibitemOpen
  \bibfield  {author} {\bibinfo {author} {\bibfnamefont {W.~L.}\ \bibnamefont
  {Faust}}\ and\ \bibinfo {author} {\bibfnamefont {L.~Y.}\ \bibnamefont
  {Chow~Chiu}},\ }\href {\doibase 10.1103/PhysRev.129.1214} {\bibfield
  {journal} {\bibinfo  {journal} {Phys. Rev.}\ }\textbf {\bibinfo {volume}
  {129}},\ \bibinfo {pages} {1214} (\bibinfo {year} {1963})}\BibitemShut
  {NoStop}%
\bibitem [{\citenamefont {Brown}\ and\ \citenamefont {King}(1966)}]{Brown1966}%
  \BibitemOpen
  \bibfield  {author} {\bibinfo {author} {\bibfnamefont {H.~H.}\ \bibnamefont
  {Brown}}\ and\ \bibinfo {author} {\bibfnamefont {J.~G.}\ \bibnamefont
  {King}},\ }\href {\doibase 10.1103/PhysRev.142.53} {\bibfield  {journal}
  {\bibinfo  {journal} {Phys. Rev.}\ }\textbf {\bibinfo {volume} {142}},\
  \bibinfo {pages} {53} (\bibinfo {year} {1966})}\BibitemShut {NoStop}%
\bibitem [{\citenamefont {Blachman}\ \emph {et~al.}(1967)\citenamefont
  {Blachman}, \citenamefont {Landman},\ and\ \citenamefont
  {Lurio}}]{Blachman1967}%
  \BibitemOpen
  \bibfield  {author} {\bibinfo {author} {\bibfnamefont {A.~G.}\ \bibnamefont
  {Blachman}}, \bibinfo {author} {\bibfnamefont {D.~A.}\ \bibnamefont
  {Landman}}, \ and\ \bibinfo {author} {\bibfnamefont {A.}~\bibnamefont
  {Lurio}},\ }\href {\doibase 10.1103/PhysRev.161.60} {\bibfield  {journal}
  {\bibinfo  {journal} {Phys. Rev.}\ }\textbf {\bibinfo {volume} {161}},\
  \bibinfo {pages} {60} (\bibinfo {year} {1967})}\BibitemShut {NoStop}%
\bibitem [{\citenamefont {Unsworth}(1969)}]{Unsworth1969}%
  \BibitemOpen
  \bibfield  {author} {\bibinfo {author} {\bibfnamefont {P.~J.}\ \bibnamefont
  {Unsworth}},\ }\href {\doibase 10.1088/0022-3700/2/1/318} {\bibfield
  {journal} {\bibinfo  {journal} {Journal of Physics B: Atomic and Molecular
  Physics}\ }\textbf {\bibinfo {volume} {2}},\ \bibinfo {pages} {122} (\bibinfo
  {year} {1969})}\BibitemShut {NoStop}%
\bibitem [{\citenamefont {Hull}\ and\ \citenamefont {Brink}(1970)}]{hull1970}%
  \BibitemOpen
  \bibfield  {author} {\bibinfo {author} {\bibfnamefont {R.~J.}\ \bibnamefont
  {Hull}}\ and\ \bibinfo {author} {\bibfnamefont {G.~O.}\ \bibnamefont
  {Brink}},\ }\href@noop {} {\bibfield  {journal} {\bibinfo  {journal}
  {Physical Review A}\ }\textbf {\bibinfo {volume} {1}},\ \bibinfo {pages}
  {685} (\bibinfo {year} {1970})}\BibitemShut {NoStop}%
\bibitem [{\citenamefont {Landman}\ and\ \citenamefont
  {Lurio}(1970)}]{Landman1970}%
  \BibitemOpen
  \bibfield  {author} {\bibinfo {author} {\bibfnamefont {D.~A.}\ \bibnamefont
  {Landman}}\ and\ \bibinfo {author} {\bibfnamefont {A.}~\bibnamefont
  {Lurio}},\ }\href {\doibase 10.1103/PhysRevA.1.1330} {\bibfield  {journal}
  {\bibinfo  {journal} {Phys. Rev. A}\ }\textbf {\bibinfo {volume} {1}},\
  \bibinfo {pages} {1330} (\bibinfo {year} {1970})}\BibitemShut {NoStop}%
\bibitem [{\citenamefont {Childs}\ \emph {et~al.}(1979)\citenamefont {Childs},
  \citenamefont {Poulsen}, \citenamefont {Goodman},\ and\ \citenamefont
  {Crosswhite}}]{Childs1978}%
  \BibitemOpen
  \bibfield  {author} {\bibinfo {author} {\bibfnamefont {W.~J.}\ \bibnamefont
  {Childs}}, \bibinfo {author} {\bibfnamefont {O.}~\bibnamefont {Poulsen}},
  \bibinfo {author} {\bibfnamefont {L.~S.}\ \bibnamefont {Goodman}}, \ and\
  \bibinfo {author} {\bibfnamefont {H.}~\bibnamefont {Crosswhite}},\ }\href
  {\doibase 10.1103/PhysRevA.19.168} {\bibfield  {journal} {\bibinfo  {journal}
  {Phys. Rev. A}\ }\textbf {\bibinfo {volume} {19}},\ \bibinfo {pages} {168}
  (\bibinfo {year} {1979})}\BibitemShut {NoStop}%
\bibitem [{\citenamefont {Brenner}\ \emph {et~al.}(1985)\citenamefont
  {Brenner}, \citenamefont {B{\"u}ttgenbach}, \citenamefont {Rupprecht},\ and\
  \citenamefont {Tr{\"a}ber}}]{Brenner1985}%
  \BibitemOpen
  \bibfield  {author} {\bibinfo {author} {\bibfnamefont {T.}~\bibnamefont
  {Brenner}}, \bibinfo {author} {\bibfnamefont {S.}~\bibnamefont
  {B{\"u}ttgenbach}}, \bibinfo {author} {\bibfnamefont {W.}~\bibnamefont
  {Rupprecht}}, \ and\ \bibinfo {author} {\bibfnamefont {F.}~\bibnamefont
  {Tr{\"a}ber}},\ }\href {\doibase
  https://doi.org/10.1016/0375-9474(85)90237-4} {\bibfield  {journal} {\bibinfo
   {journal} {Nuclear Physics A}\ }\textbf {\bibinfo {volume} {440}},\ \bibinfo
  {pages} {407 } (\bibinfo {year} {1985})}\BibitemShut {NoStop}%
\bibitem [{\citenamefont {Childs}(1991)}]{Childs1991}%
  \BibitemOpen
  \bibfield  {author} {\bibinfo {author} {\bibfnamefont {W.~J.}\ \bibnamefont
  {Childs}},\ }\href {\doibase 10.1103/PhysRevA.44.1523} {\bibfield  {journal}
  {\bibinfo  {journal} {Phys. Rev. A}\ }\textbf {\bibinfo {volume} {44}},\
  \bibinfo {pages} {1523} (\bibinfo {year} {1991})}\BibitemShut {NoStop}%
\bibitem [{\citenamefont {Jin}\ \emph {et~al.}(1995)\citenamefont {Jin},
  \citenamefont {Wakasugi}, \citenamefont {Inamura}, \citenamefont {Murayama},
  \citenamefont {Wakui}, \citenamefont {Katsuragawa}, \citenamefont {Ariga},
  \citenamefont {Ishizuka},\ and\ \citenamefont {Sugai}}]{Jin1995}%
  \BibitemOpen
  \bibfield  {author} {\bibinfo {author} {\bibfnamefont {W.~G.}\ \bibnamefont
  {Jin}}, \bibinfo {author} {\bibfnamefont {M.}~\bibnamefont {Wakasugi}},
  \bibinfo {author} {\bibfnamefont {T.~T.}\ \bibnamefont {Inamura}}, \bibinfo
  {author} {\bibfnamefont {T.}~\bibnamefont {Murayama}}, \bibinfo {author}
  {\bibfnamefont {T.}~\bibnamefont {Wakui}}, \bibinfo {author} {\bibfnamefont
  {H.}~\bibnamefont {Katsuragawa}}, \bibinfo {author} {\bibfnamefont
  {T.}~\bibnamefont {Ariga}}, \bibinfo {author} {\bibfnamefont
  {T.}~\bibnamefont {Ishizuka}}, \ and\ \bibinfo {author} {\bibfnamefont
  {I.}~\bibnamefont {Sugai}},\ }\href {\doibase 10.1103/PhysRevA.52.157}
  {\bibfield  {journal} {\bibinfo  {journal} {Phys. Rev. A}\ }\textbf {\bibinfo
  {volume} {52}},\ \bibinfo {pages} {157} (\bibinfo {year} {1995})}\BibitemShut
  {NoStop}%
\bibitem [{\citenamefont {Gerginov}\ \emph {et~al.}(2003)\citenamefont
  {Gerginov}, \citenamefont {Derevianko},\ and\ \citenamefont
  {Tanner}}]{Gerginov2003}%
  \BibitemOpen
  \bibfield  {author} {\bibinfo {author} {\bibfnamefont {V.}~\bibnamefont
  {Gerginov}}, \bibinfo {author} {\bibfnamefont {A.}~\bibnamefont
  {Derevianko}}, \ and\ \bibinfo {author} {\bibfnamefont {C.~E.}\ \bibnamefont
  {Tanner}},\ }\href {\doibase 10.1103/PhysRevLett.91.072501} {\bibfield
  {journal} {\bibinfo  {journal} {Phys. Rev. Lett.}\ }\textbf {\bibinfo
  {volume} {91}},\ \bibinfo {pages} {072501} (\bibinfo {year}
  {2003})}\BibitemShut {NoStop}%
\bibitem [{\citenamefont {Lewty}\ \emph {et~al.}(2012)\citenamefont {Lewty},
  \citenamefont {Chuah}, \citenamefont {Cazan}, \citenamefont {Sahoo},\ and\
  \citenamefont {Barrett}}]{Lewty2012}%
  \BibitemOpen
  \bibfield  {author} {\bibinfo {author} {\bibfnamefont {N.~C.}\ \bibnamefont
  {Lewty}}, \bibinfo {author} {\bibfnamefont {B.~L.}\ \bibnamefont {Chuah}},
  \bibinfo {author} {\bibfnamefont {R.}~\bibnamefont {Cazan}}, \bibinfo
  {author} {\bibfnamefont {B.~K.}\ \bibnamefont {Sahoo}}, \ and\ \bibinfo
  {author} {\bibfnamefont {M.~D.}\ \bibnamefont {Barrett}},\ }\href {\doibase
  10.1364/OE.20.021379} {\bibfield  {journal} {\bibinfo  {journal} {Opt.
  Express}\ }\textbf {\bibinfo {volume} {20}},\ \bibinfo {pages} {21379}
  (\bibinfo {year} {2012})}\BibitemShut {NoStop}%
\bibitem [{\citenamefont {Singh}\ \emph {et~al.}(2013)\citenamefont {Singh},
  \citenamefont {Angom},\ and\ \citenamefont {Natarajan}}]{Singh2013}%
  \BibitemOpen
  \bibfield  {author} {\bibinfo {author} {\bibfnamefont {A.~K.}\ \bibnamefont
  {Singh}}, \bibinfo {author} {\bibfnamefont {D.}~\bibnamefont {Angom}}, \ and\
  \bibinfo {author} {\bibfnamefont {V.}~\bibnamefont {Natarajan}},\ }\href
  {\doibase 10.1103/PhysRevA.87.012512} {\bibfield  {journal} {\bibinfo
  {journal} {Phys. Rev. A}\ }\textbf {\bibinfo {volume} {87}},\ \bibinfo
  {pages} {012512} (\bibinfo {year} {2013})}\BibitemShut {NoStop}%
\bibitem [{\citenamefont {Xiao}\ \emph {et~al.}(2020)\citenamefont {Xiao},
  \citenamefont {Li}, \citenamefont {Campbell}, \citenamefont {Dellaert},
  \citenamefont {McMillin}, \citenamefont {Ransford}, \citenamefont {Roman},
  \citenamefont {Derevianko} \emph {et~al.}}]{xiao2020}%
  \BibitemOpen
  \bibfield  {author} {\bibinfo {author} {\bibfnamefont {D.}~\bibnamefont
  {Xiao}}, \bibinfo {author} {\bibfnamefont {J.}~\bibnamefont {Li}}, \bibinfo
  {author} {\bibfnamefont {W.~C.}\ \bibnamefont {Campbell}}, \bibinfo {author}
  {\bibfnamefont {T.}~\bibnamefont {Dellaert}}, \bibinfo {author}
  {\bibfnamefont {P.}~\bibnamefont {McMillin}}, \bibinfo {author}
  {\bibfnamefont {A.}~\bibnamefont {Ransford}}, \bibinfo {author}
  {\bibfnamefont {C.}~\bibnamefont {Roman}}, \bibinfo {author} {\bibfnamefont
  {A.}~\bibnamefont {Derevianko}},  \emph {et~al.},\ }\href@noop {} {\bibfield
  {journal} {\bibinfo  {journal} {Physical Review A}\ }\textbf {\bibinfo
  {volume} {102}},\ \bibinfo {pages} {022810} (\bibinfo {year}
  {2020})}\BibitemShut {NoStop}%
\bibitem [{\citenamefont {Brink}\ and\ \citenamefont
  {Satchler}(1968)}]{brink1968}%
  \BibitemOpen
  \bibfield  {author} {\bibinfo {author} {\bibfnamefont {D.}~\bibnamefont
  {Brink}}\ and\ \bibinfo {author} {\bibfnamefont {G.}~\bibnamefont
  {Satchler}},\ }\href@noop {} {\enquote {\bibinfo {title} {Angular momentum
  2nd edn, clarendon},}\ } (\bibinfo {year} {1968})\BibitemShut {NoStop}%
\bibitem [{\citenamefont {Grant}(2007)}]{Grant2007}%
  \BibitemOpen
  \bibfield  {author} {\bibinfo {author} {\bibfnamefont {I.~P.}\ \bibnamefont
  {Grant}},\ }\href {\doibase 10.1007/978-0-387-35069-1} {\emph {\bibinfo
  {title} {{Relativistic Quantum Theory of Atoms and Molecules}}}}\ (\bibinfo
  {publisher} {Springer New York},\ \bibinfo {year} {2007})\BibitemShut
  {NoStop}%
\bibitem [{\citenamefont {{Froese Fischer}}\ \emph {et~al.}(2016)\citenamefont
  {{Froese Fischer}}, \citenamefont {Godefroid}, \citenamefont {Brage},
  \citenamefont {J{\"{o}}nsson},\ and\ \citenamefont
  {Gaigalas}}]{FroeseFischer2016a}%
  \BibitemOpen
  \bibfield  {author} {\bibinfo {author} {\bibfnamefont {C.}~\bibnamefont
  {{Froese Fischer}}}, \bibinfo {author} {\bibfnamefont {M.}~\bibnamefont
  {Godefroid}}, \bibinfo {author} {\bibfnamefont {T.}~\bibnamefont {Brage}},
  \bibinfo {author} {\bibfnamefont {P.}~\bibnamefont {J{\"{o}}nsson}}, \ and\
  \bibinfo {author} {\bibfnamefont {G.}~\bibnamefont {Gaigalas}},\ }\href
  {\doibase 10.1088/0953-4075/49/18/182004} {\bibfield  {journal} {\bibinfo
  {journal} {Journal of Physics B: Atomic, Molecular and Optical Physics}\
  }\textbf {\bibinfo {volume} {49}},\ \bibinfo {pages} {182004} (\bibinfo
  {year} {2016})}\BibitemShut {NoStop}%
\bibitem [{\citenamefont {{Froese Fischer}}\ \emph {et~al.}(2019)\citenamefont
  {{Froese Fischer}}, \citenamefont {Gaigalas}, \citenamefont {J{\"{o}}nsson},\
  and\ \citenamefont {Biero{\'{n}}}}]{FroeseFischer2019}%
  \BibitemOpen
  \bibfield  {author} {\bibinfo {author} {\bibfnamefont {C.}~\bibnamefont
  {{Froese Fischer}}}, \bibinfo {author} {\bibfnamefont {G.}~\bibnamefont
  {Gaigalas}}, \bibinfo {author} {\bibfnamefont {P.}~\bibnamefont
  {J{\"{o}}nsson}}, \ and\ \bibinfo {author} {\bibfnamefont {J.}~\bibnamefont
  {Biero{\'{n}}}},\ }\href {\doibase 10.1016/j.cpc.2018.10.032} {\bibfield
  {journal} {\bibinfo  {journal} {Computer Physics Communications}\ }\textbf
  {\bibinfo {volume} {237}},\ \bibinfo {pages} {184} (\bibinfo {year}
  {2019})}\BibitemShut {NoStop}%
\bibitem [{\citenamefont {Li}\ \emph {et~al.}()\citenamefont {Li},
  \citenamefont {Ekman}, \citenamefont {Gediminas}, \citenamefont
  {Biero{\'{n}}}, \citenamefont {J{\"{o}}nsson}, \citenamefont {Godefroid},\
  and\ \citenamefont {{Froese Fischer}}}]{JGLiunpublished}%
  \BibitemOpen
  \bibfield  {author} {\bibinfo {author} {\bibfnamefont {J.}~\bibnamefont
  {Li}}, \bibinfo {author} {\bibfnamefont {J.}~\bibnamefont {Ekman}}, \bibinfo
  {author} {\bibfnamefont {G.}~\bibnamefont {Gediminas}}, \bibinfo {author}
  {\bibfnamefont {J.}~\bibnamefont {Biero{\'{n}}}}, \bibinfo {author}
  {\bibfnamefont {P.}~\bibnamefont {J{\"{o}}nsson}}, \bibinfo {author}
  {\bibfnamefont {M.}~\bibnamefont {Godefroid}}, \ and\ \bibinfo {author}
  {\bibfnamefont {C.}~\bibnamefont {{Froese Fischer}}},\ }\href@noop {}
  {\bibinfo  {journal} {Computer Physics Communications (in preparation)}\
  }\BibitemShut {NoStop}%
\bibitem [{\citenamefont {J{\"{o}}nsson}\ \emph {et~al.}(1996)\citenamefont
  {J{\"{o}}nsson}, \citenamefont {Parpia},\ and\ \citenamefont {{Froese
  Fischer}}}]{Jonsson1996}%
  \BibitemOpen
\bibfield  {journal} {  }\bibfield  {author} {\bibinfo {author} {\bibfnamefont
  {P.}~\bibnamefont {J{\"{o}}nsson}}, \bibinfo {author} {\bibfnamefont
  {F.}~\bibnamefont {Parpia}}, \ and\ \bibinfo {author} {\bibfnamefont
  {C.}~\bibnamefont {{Froese Fischer}}},\ }\href@noop {} {\bibfield  {journal}
  {\bibinfo  {journal} {Computer Physics Communications}\ }\textbf {\bibinfo
  {volume} {96}},\ \bibinfo {pages} {301} (\bibinfo {year} {1996})}\BibitemShut
  {NoStop}%
\bibitem [{\citenamefont {Li}\ \emph {et~al.}(2012)\citenamefont {Li},
  \citenamefont {J{\"{o}}nsson}, \citenamefont {Godefroid}, \citenamefont
  {Dong},\ and\ \citenamefont {Gaigalas}}]{Li2012}%
  \BibitemOpen
  \bibfield  {author} {\bibinfo {author} {\bibfnamefont {J.~G.}\ \bibnamefont
  {Li}}, \bibinfo {author} {\bibfnamefont {P.}~\bibnamefont {J{\"{o}}nsson}},
  \bibinfo {author} {\bibfnamefont {M.}~\bibnamefont {Godefroid}}, \bibinfo
  {author} {\bibfnamefont {C.~Z.}\ \bibnamefont {Dong}}, \ and\ \bibinfo
  {author} {\bibfnamefont {G.}~\bibnamefont {Gaigalas}},\ }\href {\doibase
  10.1103/PhysRevA.86.052523} {\bibfield  {journal} {\bibinfo  {journal}
  {Physical Review A}\ }\textbf {\bibinfo {volume} {86}},\ \bibinfo {pages}
  {052523} (\bibinfo {year} {2012})}\BibitemShut {NoStop}%
\bibitem [{\citenamefont {Li}\ \emph {et~al.}(2016)\citenamefont {Li},
  \citenamefont {Godefroid},\ and\ \citenamefont {Wang}}]{Li2016a}%
  \BibitemOpen
  \bibfield  {author} {\bibinfo {author} {\bibfnamefont {J.}~\bibnamefont
  {Li}}, \bibinfo {author} {\bibfnamefont {M.}~\bibnamefont {Godefroid}}, \
  and\ \bibinfo {author} {\bibfnamefont {J.}~\bibnamefont {Wang}},\ }\href
  {\doibase 10.1088/0953-4075/49/11/115002} {\bibfield  {journal} {\bibinfo
  {journal} {Journal of Physics B: Atomic, Molecular and Optical Physics}\
  }\textbf {\bibinfo {volume} {49}},\ \bibinfo {pages} {115002} (\bibinfo
  {year} {2016})}\BibitemShut {NoStop}%
\bibitem [{\citenamefont {Porsev}\ \emph {et~al.}(1999)\citenamefont {Porsev},
  \citenamefont {Rakhlina},\ and\ \citenamefont {Kozlov}}]{Porsev1999}%
  \BibitemOpen
  \bibfield  {author} {\bibinfo {author} {\bibfnamefont {S.}~\bibnamefont
  {Porsev}}, \bibinfo {author} {\bibfnamefont {Y.~G.}\ \bibnamefont
  {Rakhlina}}, \ and\ \bibinfo {author} {\bibfnamefont {M.}~\bibnamefont
  {Kozlov}},\ }\href@noop {} {\bibfield  {journal} {\bibinfo  {journal}
  {Journal of Physics B: Atomic, Molecular and Optical Physics}\ }\textbf
  {\bibinfo {volume} {32}},\ \bibinfo {pages} {1113} (\bibinfo {year}
  {1999})}\BibitemShut {NoStop}%
\bibitem [{\citenamefont {Stone}(2005)}]{Stone2005}%
  \BibitemOpen
  \bibfield  {author} {\bibinfo {author} {\bibfnamefont {N.~J.}\ \bibnamefont
  {Stone}},\ }\href {\doibase 10.1016/j.adt.2005.04.001} {\bibfield  {journal}
  {\bibinfo  {journal} {Atomic Data and Nuclear Data Tables}\ }\textbf
  {\bibinfo {volume} {90}},\ \bibinfo {pages} {75} (\bibinfo {year}
  {2005})}\BibitemShut {NoStop}%
\bibitem [{\citenamefont {Bohr}\ and\ \citenamefont
  {Weisskopf}(1950)}]{Bohr1950}%
  \BibitemOpen
  \bibfield  {author} {\bibinfo {author} {\bibfnamefont {A.}~\bibnamefont
  {Bohr}}\ and\ \bibinfo {author} {\bibfnamefont {V.~F.}\ \bibnamefont
  {Weisskopf}},\ }\href@noop {} {\bibfield  {journal} {\bibinfo  {journal}
  {Phys. Rev.}\ }\textbf {\bibinfo {volume} {77}},\ \bibinfo {pages} {94}
  (\bibinfo {year} {1950})}\BibitemShut {NoStop}%
\bibitem [{\citenamefont {Kramida}\ \emph {et~al.}(2019)\citenamefont
  {Kramida}, \citenamefont {Ralchenko}, \citenamefont {Reader},\ and\
  \citenamefont {{NIST ASD TEAM}}}]{NIST}%
  \BibitemOpen
  \bibfield  {author} {\bibinfo {author} {\bibfnamefont {A.}~\bibnamefont
  {Kramida}}, \bibinfo {author} {\bibfnamefont {Y.}~\bibnamefont {Ralchenko}},
  \bibinfo {author} {\bibfnamefont {J.}~\bibnamefont {Reader}}, \ and\ \bibinfo
  {author} {\bibnamefont {{NIST ASD TEAM}}},\ }\href {\doibase
  https://doi.org/10.18434/T4W30F} {\enquote {\bibinfo {title} {{NIST Atomic
  Spectra Database (version 5.7) [Online]}},}\ } (\bibinfo {year}
  {2019})\BibitemShut {NoStop}%
\bibitem [{\citenamefont {Vernon}\ \emph {et~al.}(2019)\citenamefont {Vernon},
  \citenamefont {Billowes}, \citenamefont {Binnersley}, \citenamefont
  {Bissell}, \citenamefont {Cocolios}, \citenamefont {Farooq-Smith},
  \citenamefont {Flanagan}, \citenamefont {Ruiz}, \citenamefont {Gins},
  \citenamefont {de~Groote}, \citenamefont {Koszor{\'u}s}, \citenamefont
  {Lynch}, \citenamefont {Neyens}, \citenamefont {Ricketts}, \citenamefont
  {Wendt}, \citenamefont {Wilkins},\ and\ \citenamefont {Yang}}]{Vernon2019}%
  \BibitemOpen
  \bibfield  {author} {\bibinfo {author} {\bibfnamefont {A.}~\bibnamefont
  {Vernon}}, \bibinfo {author} {\bibfnamefont {J.}~\bibnamefont {Billowes}},
  \bibinfo {author} {\bibfnamefont {C.}~\bibnamefont {Binnersley}}, \bibinfo
  {author} {\bibfnamefont {M.}~\bibnamefont {Bissell}}, \bibinfo {author}
  {\bibfnamefont {T.}~\bibnamefont {Cocolios}}, \bibinfo {author}
  {\bibfnamefont {G.}~\bibnamefont {Farooq-Smith}}, \bibinfo {author}
  {\bibfnamefont {K.}~\bibnamefont {Flanagan}}, \bibinfo {author}
  {\bibfnamefont {R.~G.}\ \bibnamefont {Ruiz}}, \bibinfo {author}
  {\bibfnamefont {W.}~\bibnamefont {Gins}}, \bibinfo {author} {\bibfnamefont
  {R.}~\bibnamefont {de~Groote}}, \bibinfo {author} {\bibfnamefont
  {{\'A}.}~\bibnamefont {Koszor{\'u}s}}, \bibinfo {author} {\bibfnamefont
  {K.}~\bibnamefont {Lynch}}, \bibinfo {author} {\bibfnamefont
  {G.}~\bibnamefont {Neyens}}, \bibinfo {author} {\bibfnamefont
  {C.}~\bibnamefont {Ricketts}}, \bibinfo {author} {\bibfnamefont
  {K.}~\bibnamefont {Wendt}}, \bibinfo {author} {\bibfnamefont
  {S.}~\bibnamefont {Wilkins}}, \ and\ \bibinfo {author} {\bibfnamefont
  {X.}~\bibnamefont {Yang}},\ }\href {\doibase
  https://doi.org/10.1016/j.sab.2019.02.001} {\bibfield  {journal} {\bibinfo
  {journal} {Spectrochimica Acta Part B: Atomic Spectroscopy}\ }\textbf
  {\bibinfo {volume} {153}},\ \bibinfo {pages} {61 } (\bibinfo {year}
  {2019})}\BibitemShut {NoStop}%
\bibitem [{\citenamefont {Vormawah}\ \emph {et~al.}(2018)\citenamefont
  {Vormawah}, \citenamefont {Vil{\'e}n}, \citenamefont {Beerwerth},
  \citenamefont {Campbell}, \citenamefont {Cheal}, \citenamefont {Dicker},
  \citenamefont {Eronen}, \citenamefont {Fritzsche}, \citenamefont {Geldhof},
  \citenamefont {Jokinen} \emph {et~al.}}]{vormawah2018}%
  \BibitemOpen
  \bibfield  {author} {\bibinfo {author} {\bibfnamefont {L.}~\bibnamefont
  {Vormawah}}, \bibinfo {author} {\bibfnamefont {M.}~\bibnamefont {Vil{\'e}n}},
  \bibinfo {author} {\bibfnamefont {R.}~\bibnamefont {Beerwerth}}, \bibinfo
  {author} {\bibfnamefont {P.}~\bibnamefont {Campbell}}, \bibinfo {author}
  {\bibfnamefont {B.}~\bibnamefont {Cheal}}, \bibinfo {author} {\bibfnamefont
  {A.}~\bibnamefont {Dicker}}, \bibinfo {author} {\bibfnamefont
  {T.}~\bibnamefont {Eronen}}, \bibinfo {author} {\bibfnamefont
  {S.}~\bibnamefont {Fritzsche}}, \bibinfo {author} {\bibfnamefont
  {S.}~\bibnamefont {Geldhof}}, \bibinfo {author} {\bibfnamefont
  {A.}~\bibnamefont {Jokinen}},  \emph {et~al.},\ }\href@noop {} {\bibfield
  {journal} {\bibinfo  {journal} {Physical Review A}\ }\textbf {\bibinfo
  {volume} {97}},\ \bibinfo {pages} {042504} (\bibinfo {year}
  {2018})}\BibitemShut {NoStop}%
\bibitem [{\citenamefont {Gins}\ \emph {et~al.}(2018)\citenamefont {Gins},
  \citenamefont {de~Groote}, \citenamefont {Bissell}, \citenamefont {Buitrago},
  \citenamefont {Ferrer}, \citenamefont {Lynch}, \citenamefont {Neyens},\ and\
  \citenamefont {Sels}}]{gins2018}%
  \BibitemOpen
  \bibfield  {author} {\bibinfo {author} {\bibfnamefont {W.}~\bibnamefont
  {Gins}}, \bibinfo {author} {\bibfnamefont {R.~P.}\ \bibnamefont {de~Groote}},
  \bibinfo {author} {\bibfnamefont {M.~L.}\ \bibnamefont {Bissell}}, \bibinfo
  {author} {\bibfnamefont {C.~G.}\ \bibnamefont {Buitrago}}, \bibinfo {author}
  {\bibfnamefont {R.}~\bibnamefont {Ferrer}}, \bibinfo {author} {\bibfnamefont
  {K.~M.}\ \bibnamefont {Lynch}}, \bibinfo {author} {\bibfnamefont
  {G.}~\bibnamefont {Neyens}}, \ and\ \bibinfo {author} {\bibfnamefont
  {S.}~\bibnamefont {Sels}},\ }\href
  {https://doi.org/10.1016/j.cpc.2017.09.012} {\bibfield  {journal} {\bibinfo
  {journal} {Comput. Phys. Comm.}\ }\textbf {\bibinfo {volume} {222}},\
  \bibinfo {pages} {286} (\bibinfo {year} {2018})}\BibitemShut {NoStop}%
\bibitem [{\citenamefont {de~Groote}\ \emph {et~al.}(2020)\citenamefont
  {de~Groote}, \citenamefont {de~Roubin}, \citenamefont {Campbell},
  \citenamefont {Cheal}, \citenamefont {Devlin}, \citenamefont {Eronen},
  \citenamefont {Geldhof}, \citenamefont {Moore}, \citenamefont {Reponen},
  \citenamefont {Rinta-Antila},\ and\ \citenamefont {Schuh}}]{Degroote2019}%
  \BibitemOpen
  \bibfield  {author} {\bibinfo {author} {\bibfnamefont {R.}~\bibnamefont
  {de~Groote}}, \bibinfo {author} {\bibfnamefont {A.}~\bibnamefont
  {de~Roubin}}, \bibinfo {author} {\bibfnamefont {P.}~\bibnamefont {Campbell}},
  \bibinfo {author} {\bibfnamefont {B.}~\bibnamefont {Cheal}}, \bibinfo
  {author} {\bibfnamefont {C.}~\bibnamefont {Devlin}}, \bibinfo {author}
  {\bibfnamefont {T.}~\bibnamefont {Eronen}}, \bibinfo {author} {\bibfnamefont
  {S.}~\bibnamefont {Geldhof}}, \bibinfo {author} {\bibfnamefont
  {I.}~\bibnamefont {Moore}}, \bibinfo {author} {\bibfnamefont
  {M.}~\bibnamefont {Reponen}}, \bibinfo {author} {\bibfnamefont
  {S.}~\bibnamefont {Rinta-Antila}}, \ and\ \bibinfo {author} {\bibfnamefont
  {M.}~\bibnamefont {Schuh}},\ }\href {\doibase
  https://doi.org/10.1016/j.nimb.2019.04.028} {\bibfield  {journal} {\bibinfo
  {journal} {Nuclear Instruments and Methods in Physics Research Section B:
  Beam Interactions with Materials and Atoms}\ }\textbf {\bibinfo {volume}
  {463}},\ \bibinfo {pages} {437 } (\bibinfo {year} {2020})}\BibitemShut
  {NoStop}%
\bibitem [{\citenamefont {Angeli}\ and\ \citenamefont
  {Marinova}(2013)}]{Angeli2013}%
  \BibitemOpen
  \bibfield  {author} {\bibinfo {author} {\bibfnamefont {I.}~\bibnamefont
  {Angeli}}\ and\ \bibinfo {author} {\bibfnamefont {K.}~\bibnamefont
  {Marinova}},\ }\href {\doibase https://doi.org/10.1016/j.adt.2011.12.006}
  {\bibfield  {journal} {\bibinfo  {journal} {Atomic Data and Nuclear Data
  Tables}\ }\textbf {\bibinfo {volume} {99}},\ \bibinfo {pages} {69 } (\bibinfo
  {year} {2013})}\BibitemShut {NoStop}%
\bibitem [{\citenamefont {Schwartz}(1955)}]{Schwartz1955}%
  \BibitemOpen
  \bibfield  {author} {\bibinfo {author} {\bibfnamefont {C.}~\bibnamefont
  {Schwartz}},\ }\href {\doibase 10.1103/PhysRev.97.380} {\bibfield  {journal}
  {\bibinfo  {journal} {Phys. Rev.}\ }\textbf {\bibinfo {volume} {97}},\
  \bibinfo {pages} {380} (\bibinfo {year} {1955})}\BibitemShut {NoStop}%
\bibitem [{\citenamefont {Williams}(1962)}]{Williams1962}%
  \BibitemOpen
  \bibfield  {author} {\bibinfo {author} {\bibfnamefont {S.~A.}\ \bibnamefont
  {Williams}},\ }\href {\doibase 10.1103/PhysRev.125.340} {\bibfield  {journal}
  {\bibinfo  {journal} {Phys. Rev.}\ }\textbf {\bibinfo {volume} {125}},\
  \bibinfo {pages} {340} (\bibinfo {year} {1962})}\BibitemShut {NoStop}%
\bibitem [{\citenamefont {Brown}\ \emph {et~al.}(1980)\citenamefont {Brown},
  \citenamefont {Chung},\ and\ \citenamefont {Wildenthal}}]{brown1980}%
  \BibitemOpen
  \bibfield  {author} {\bibinfo {author} {\bibfnamefont {B.~A.}\ \bibnamefont
  {Brown}}, \bibinfo {author} {\bibfnamefont {W.}~\bibnamefont {Chung}}, \ and\
  \bibinfo {author} {\bibfnamefont {B.}~\bibnamefont {Wildenthal}},\
  }\href@noop {} {\bibfield  {journal} {\bibinfo  {journal} {Phys. Rev. C}\
  }\textbf {\bibinfo {volume} {22}},\ \bibinfo {pages} {774} (\bibinfo {year}
  {1980})}\BibitemShut {NoStop}%
\bibitem [{\citenamefont {Amoruso}\ and\ \citenamefont
  {Johnson}(1971)}]{Amoruso1971}%
  \BibitemOpen
  \bibfield  {author} {\bibinfo {author} {\bibfnamefont {M.~J.}\ \bibnamefont
  {Amoruso}}\ and\ \bibinfo {author} {\bibfnamefont {W.~R.}\ \bibnamefont
  {Johnson}},\ }\href {\doibase 10.1103/PhysRevA.3.6} {\bibfield  {journal}
  {\bibinfo  {journal} {Phys. Rev. A}\ }\textbf {\bibinfo {volume} {3}},\
  \bibinfo {pages} {6} (\bibinfo {year} {1971})}\BibitemShut {NoStop}%
\bibitem [{\citenamefont {Lewty}\ \emph {et~al.}(2013)\citenamefont {Lewty},
  \citenamefont {Chuah}, \citenamefont {Cazan}, \citenamefont {Barrett},\ and\
  \citenamefont {Sahoo}}]{Lewty2013}%
  \BibitemOpen
  \bibfield  {author} {\bibinfo {author} {\bibfnamefont {N.~C.}\ \bibnamefont
  {Lewty}}, \bibinfo {author} {\bibfnamefont {B.~L.}\ \bibnamefont {Chuah}},
  \bibinfo {author} {\bibfnamefont {R.}~\bibnamefont {Cazan}}, \bibinfo
  {author} {\bibfnamefont {M.~D.}\ \bibnamefont {Barrett}}, \ and\ \bibinfo
  {author} {\bibfnamefont {B.~K.}\ \bibnamefont {Sahoo}},\ }\href {\doibase
  10.1103/PhysRevA.88.012518} {\bibfield  {journal} {\bibinfo  {journal} {Phys.
  Rev. A}\ }\textbf {\bibinfo {volume} {88}},\ \bibinfo {pages} {012518}
  (\bibinfo {year} {2013})}\BibitemShut {NoStop}%
\bibitem [{\citenamefont {Fuller}(1976)}]{fuller1976}%
  \BibitemOpen
  \bibfield  {author} {\bibinfo {author} {\bibfnamefont {G.~H.}\ \bibnamefont
  {Fuller}},\ }\href@noop {} {\bibfield  {journal} {\bibinfo  {journal}
  {Journal of Physical and Chemical Reference Data}\ }\textbf {\bibinfo
  {volume} {5}},\ \bibinfo {pages} {835} (\bibinfo {year} {1976})}\BibitemShut
  {NoStop}%
\end{thebibliography}
\end{document}